\documentclass[preprint,prd,amsmath,amssymb,axodraw,showpacs]{revtex4}

\usepackage{epsfig}
\usepackage{dcolumn}
\usepackage{bm}
\usepackage{amssymb, amsmath, amsthm}
\usepackage{graphicx}

\begin{document}



\title{Gluon fragmentation functions in the Nambu-Jona-Lasinio model}


\author{Dong-Jing Yang}\email{djyang@std.ntnu.edu.tw}
\affiliation{Department of Physics, National Taiwan Normal University,
Taipei 10610, Taiwan, Republic of China}
\author{Hsiang-nan Li}\email{hnli@phys.sinica.edu.tw}
\affiliation{Institute of Physics, Academia Sinica, Taipei 11529, Taiwan, Republic of China}


\date{\today}

\begin{abstract}
We derive gluon fragmentation functions in the Nambu-Jona-Lasinio (NJL) model
by treating a gluon as a pair of color lines formed by fictitious quark and
anti-quark ($q\bar q$). Gluon elementary fragmentation functions are obtained
from the quark and anti-quark elementary fragmentation functions for emitting
specific mesons in the NJL model under the requirement that the $q\bar q$
pair maintains in the flavor-singlet state after meson emissions. An integral
equation, which iterates the gluon elementary fragmentation functions to all
orders, is then solved to yield the gluon fragmentation functions at a model scale.
It is observed that these solutions are stable with respect to variation of
relevant model parameters, especially after QCD evolution to a higher scale is
implemented. We show that the inclusion of the gluon fragmentation functions
into the theoretical predictions from only the quark fragmentation functions
greatly improves the agreement with the SLD data for the pion and kaon
productions in $e^+e^-$ annihilation. Our proposal provides a plausible
construct for the gluon fragmentation functions, which are supposed to be null
in the NJL model.

\end{abstract}

\pacs{12.39.Ki,13.60.Le,13.66.Bc}

\maketitle

\section{INTRODUCTION}

A fragmentation function contains important information on the strong
dynamics of producing a hadron in high-energy scattering process. It
describes the probability of a parton to emit mesons with certain fractions
of the parent parton momentum, and serves as a crucial input to a framework
for hadron production based on factorization theorem. For example, one needs
unpolarized fragmentation functions for the analysis of semi-inclusive
deeply inelastic scattering, electron-positron annihilation into hadrons,
and hadron hadroproduction \cite{ref1,ref2,ref3,ref4,ref5,ref6,ref7,ref8,
ref9,ref10,ref11}. Quark fragmentation functions in the low energy limit have
been calculated in effective models recently, such as the Nambu-Jona-Lasinio
(NJL) model \cite{ref12} and the nonlocal chiral quark
model \cite{ref13}. A concern is that gluon fragmentation
functions are assumed to be null at a model scale, due to the absence
of gluonic degrees of freedom in the corresponding Lagrangians. A gluon
can certainly fragment into hadrons at a low scale,
just like a quark does. Without the gluon fragmentation functions at a
model scale, QCD evolution effects cannot be complete, and the resultant
quark fragmentation functions at a high scale are not reliable. A simple
argument is as follows. The quark fragmentation function $D_q^h(z)$ of a
hadron $h$ with a momentum fraction $z$ obeys the
sum rule $\sum_h{\int zD_q^h(z)dz=1}$ for 100\% quark light-cone momentum
transfer to hadrons at a model scale. Setting the gluon fragmentation
function $D_g^h(z)$ to zero will violate the sum rule $\sum_h{\int zD_g^h(z)dz=1}$,
so the quark and gluon fragmentation functions invalidate
the sum rules after QCD evolution.

In this paper we attempt to derive the gluon fragmentation functions in the
NJL model. Though there are lack of gluonic degrees of freedom, we regard
a gluon as a pair of color lines formed by fictitious quark and anti-quark
($q\bar q$) in a color-octet state. A requirement is that the $q\bar q$ pair
remains flavor-singlet after meson emissions. Namely, the quark $q$ emits a
hadron $m=q\bar Q$ and the anti-quark $\bar{q}$ emits an anti-hadron $\bar m=Q\bar q$
at the same time, resulting in a flavor-singlet fictitious $Q\bar{Q}$ pair.
The idea originates from the color dipole model, in which a gluon is treated
as a pair of color lines, and parton emissions are turned into emissions of
color dipoles composed of quarks and anti-quarks. The simplest version of
our proposal leads to the formulation of the gluon fragmentation functions
in terms of the quark fragmentation functions, similar to that in the Lund
model \cite{ref14}. A refined version is to include the mechanism of quark
annihilation, which respects the flavor-singlet requirement on the $q\bar q$
pair, such that the specific flavor of the fictitious quarks is irrelevant.
Gluon elementary fragmentation functions in the refined version
are constructed from the quark and anti-quark elementary fragmentation
functions for emitting specific mesons in one step. An integral equation, which
iterates the gluon elementary fragmentation functions to all orders, is then
solved to yield the gluon fragmentation functions.

It will be verified that our results are stable with respect to variation
of relevant model parameters, including the model scales and the fictitious quark
masses, especially after QCD evolution to a higher scale is implemented. The
possible effect from branching of a gluon into more, i.e., from the multi-dipole
contribution, is also investigated, and found to be minor. With the gluon and quark
fragmentation functions obtained in this paper, we predict the $e^++e^-\rightarrow h+X$
differential cross section at the scale $Q^2=M_Z^2$, $M_Z$ being the $Z$ boson mass,
and compare it with the measured ones, such as those from TASSO \cite{ref15,ref16,ref17},
TPC \cite{ref18}, HRS \cite{ref19}, TOPAZ \cite{ref20}, SLD \cite{ref21}, ALEPH
\cite{ref22}, OPAL \cite{ref23}, and DELPHI \cite{ref24,ref25}. Since the above data
are similar, we will focus on the SLD one. It will be demonstrated, as an
appropriate model scale is chosen, that the inclusion of the gluon fragmentation
functions into the predictions from only the quark fragmentation functions greatly
improves the agreement with the SLD data for the pion and kaon productions. This work
explores  the behavior of the gluon fragmentation functions at low energy, and their
importance on phenomenological applications.

The rest of the paper is organized as follows. We review the evaluation of the
quark fragmentation functions in the NJL model in Sec. II. The color dipole
model is briefly introduced in Sec. III, which motivates our proposal to
treat a gluon as a pair of color lines. The gluon fragmentation functions are
then formulated in the simple version, which is consistent with the Lund model,
and in the refined version, which includes the quark annihilation mechanism
and the multi-dipole contribution. Numerical results of the gluon fragmentation
functions for the pion and kaon, before and after the next-to-leading-order
(NLO) QCD evolution, are presented. These results are compared to the
Hirai-Kumano-Nagai-Sudoh (HKNS) \cite{ref26} and de Florian-Sassot-Stratmann
(DSS) \cite{ref27} parameterizations of the quark and gluon fragmentation
functions in Sec. IV, and then to the SLD data of the $e^++e^-\rightarrow h+X$
differential cross sections at $Q^2=M_Z^2$. Section V contains the conclusion. Some
numerical results from the leading-order (LO) QCD evolution are collected
in the Appendix for reference.

\section{QUARK FRAGMENTATION FUNCTIONS}

The NJL model \cite{ref28,ref29} is a low-energy effective theory, like the
BCS theory, to demonstrate the chiral symmetry breaking and appearance of
Nambu-Goldstone bosons. A non-vanishing chiral condensate would be generated
as the coupling of the four-fermion interaction is greater than a critical
value. The spontaneous chiral symmetry breaking then gives rise to
dynamical quark mass from a gap equation. The spontaneous chiral symmetry
breaking also induces massless Nambu-Goldstone bosons, represented by
the pole of the summation of fermion loops to all orders in the
four-fermion coupling, and regarded as quark-antiquark excitations
of the spontaneously broken vacuum. To get massive Nambu-Goldstone bosons,
one adds a bare fermion mass term, i.e., explicit chiral symmetry breaking,
into the effective theory. The NJL model has been applied to the calculation
of quark distribution functions \cite{ref30,ref31} and fragmentation functions
\cite{ref12} for massive pseudoscalar mesons.

\begin{figure*}
 \centering
\includegraphics[scale=0.70]{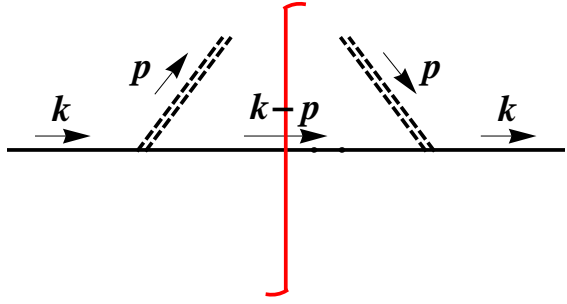}
\caption{\label{fig3}
Quark elementary fragmentation function for a pseudoscalar meson, in which the solid
and dashed lines represent the quark and the pseudoscalar meson, respectively.}
\end{figure*}

In this section we briefly review the derivation of quark fragmentation
functions for pseudoscalar mesons in the NJL model, which starts with
the construction of an elementary fragmentation function $d_q^m(z)$. This
function represents the probability of a quark $q$ to emit a meson $m$ in
one step, which carries a light-cone momentum fraction $z$ of the quark
momentum in the minus direction, as depicted in Fig.~\ref{fig3}.
In the light-cone frame the quark possesses vanishing transverse momentum
before the emission, and nonzero $k_T=-p_\perp/z$ with respect to the
direction of the emitted meson. The elementary quark fragmentation function
has been computed as \cite{ref32}
\begin{equation}
\begin{aligned}
d_q^{m}(z)=&-\frac{C_q^m}{2} g_{mqQ}^2 \frac{z}{2}
{\int}\frac{d^4k}{(2\pi)^4} tr\left[ S_1(k)\gamma^+
S_1(k)\gamma_5 (k\!\!\!/- p\!\!\!/+M_2) \gamma_5 \right]\\
&\times \delta(k_--p_-/z) 2\pi \delta((k-p)^2-M_2^2)\\
=&\frac{C_q^m}{2} g_{mqQ}^2 \frac{z}{2}{\int}\frac{d^2p_\perp}{(2\pi)^3}\frac{p_\perp^2+[(z-1)M_1-M_2]^2}
{[p_\perp^2+z(z-1)M_1^2+zM_2^2+(1-z)m_m^2]^2},\\
\end{aligned}\label{dqm}
\end{equation}
where $C_q^m$ is a flavor factor, $S_1$ denotes the quark propagator, $M_1$
and $M_2$ are the quark constituent masses before and after the emission, respectively,
and $m_m$ is the meson mass. The dipole regulator in \cite{ref33} has been employed
to avoid a divergence in the above integral. The quark-meson coupling $g_{mqQ}$ is
determined via the quark-bubble graph \cite{ref32,ref33}:
\begin{equation}
\begin{aligned}
\frac{1}{g_{mqQ}^2}&=-\frac{\partial \Pi(p)}{\partial p^2}\Big\vert_{p^2=m_m^2},\\
\Pi(p)=2 N_c i &\int \frac{d^4k}{(2\pi)^4} tr\left[\gamma_5 S_1(k) \gamma_5 S_1(k-p)\right],\\
\end{aligned}
\end{equation}
with the number of colors $N_c$.

\begin{figure*}
 \centering
\includegraphics[scale=0.75]{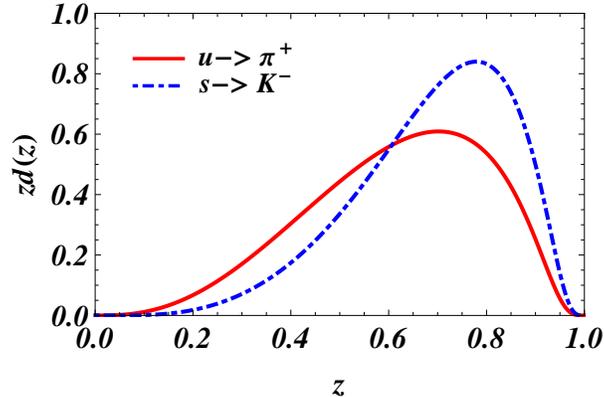}
\caption{\label{fig4}
$z$ dependence of $zd_u^{\pi^+}(z)$ and $zd_s^{K^-}(z)$.}
\end{figure*}

We adopt the values $g_{\pi qQ}=4.24$ and $g_{KqQ}=4.52$ for the couplings,
$M_u=M_d=0.4$ GeV and $M_s=0.59$ GeV for the quark constituent masses, and
$m_\pi=0.14$ GeV and $m_K=0.495$ GeV for the meson masses.
The curves of $zd_q^m(z)$ displayed in Fig.~\ref{fig4}
indicate that the probability for emitting a meson with a vanishing momentum
is tiny, and the meson which can be directly formed from the
quark $q$ in one step, such as the $u\to\pi^+$ and $s\to K^-$ channels,
prefers a momentum fraction as high as $z\sim 0.7$-0.8. Since a kaon is more
massive than a pion, it tends to carry a bit larger momentum fraction $z$.
These features will help understanding our numerical results for the gluon
fragmentation functions to be evaluated in the next section.

The integral equation based on a multiplicative ansatz for
a fragmentation function is written as \cite{ref34}
\begin{equation}
\begin{aligned}
D_q^m(z)=&\hat{d}_q^m(z)+\sum_{Q}{\int_z^1\frac{dy}{y}\hat{d}_q^Q(y)D_Q^m(\frac{z}{y})},\\
&\hat{d}_q^Q(y)=\hat{d}_q^m(1-y)|_{m=q\bar{Q}},
\end{aligned}\label{Dqm}
\end{equation}
where the elementary fragmentation function has been normalized into $\hat{d}_q^m(z)$
in order to have the meaning of
probability. Equation~(\ref{Dqm}), which iterates Eq.~(\ref{dqm}) to all orders,
determines the probability of emitting a meson $m$ by the quark $q$ with a
momentum fraction $z$ through a jet process: the first term $\hat{d}_q^m$
on the right-hand side of Eq.~(\ref{Dqm}) corresponds to the first
emission of the meson $m=q\bar Q$ in the jet process, and the second term,
containing a convolution, collects the contribution from the rest of meson
emissions in the jet process described by $D_Q^m$ with the probability
$\hat{d}_q^Q$.

\begin{figure*}
 \centering
\includegraphics[scale=0.70]{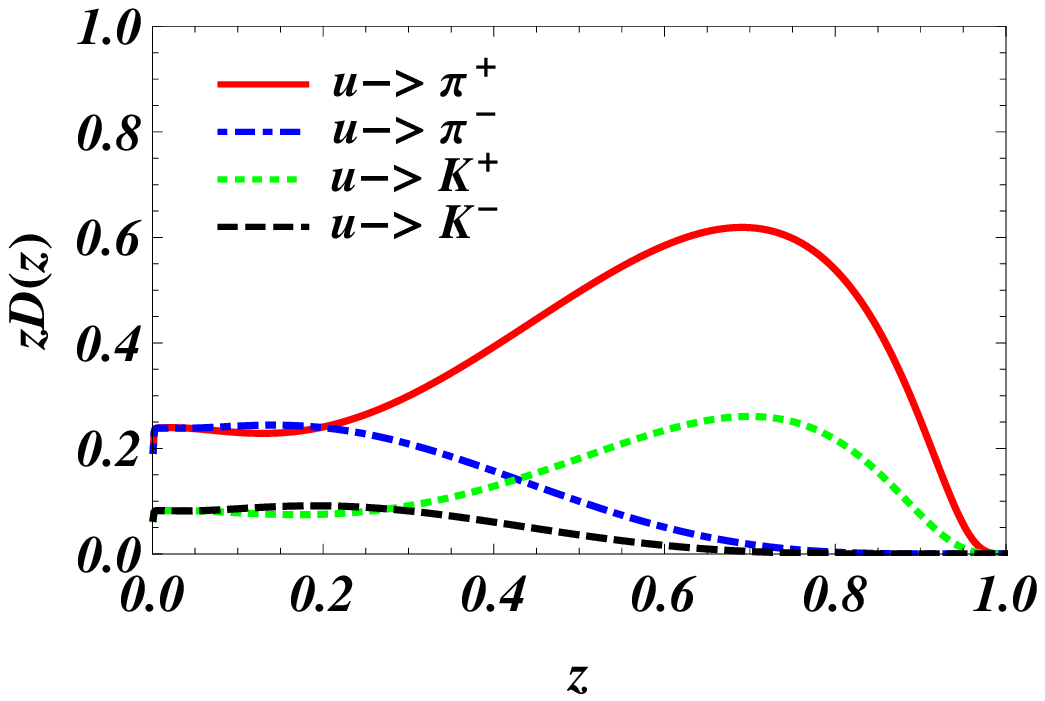}
\includegraphics[scale=0.70]{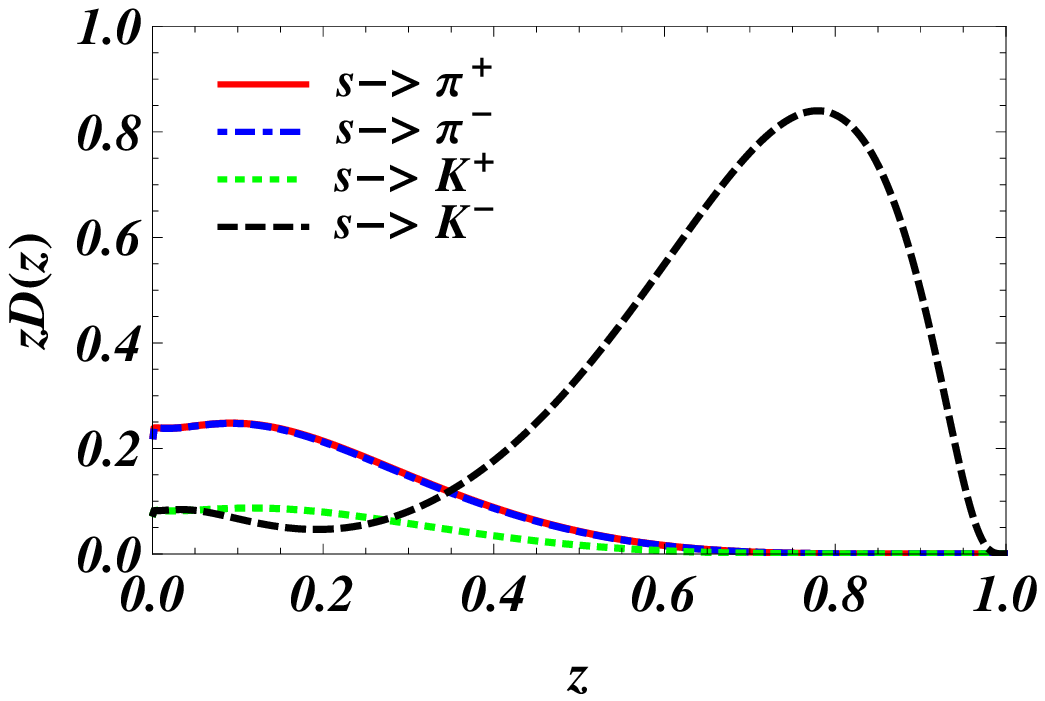}

\hspace{1.0cm}(a)\hspace{7.0cm}(b)
\caption{\label{fig5}
$z$ dependence of (a) $zD_u^m(z)$ and (b) $zD_s^m(z)$ from the NJL model at a model scale
for $m=\pi^{\pm}$ and $K^{\pm}$.}
\end{figure*}

Equation~(\ref{Dqm}) can be solved in at least three different ways to get
the quark fragmentation functions, by iteration, by inverse matrix inversion,
and by Monte Carlo simulation. Here we take the former two methods, and have
confirmed that the results are the same. The $z$ dependence of $zD_q^m(z)$ for
$q=u,s$ and $m=\pi^\pm, K^\pm$ at a model scale is exhibited in Fig.~\ref{fig5}.
It is found that the quark fragmentation functions will have peaks in the high $z$
region, if the mesons can be formed directly from the quarks  (referred to the
discussion on Fig.~\ref{fig4}), such as the $u\to\pi^+, K^+$ and $s\to K^-$ channels.
Otherwise, the mesons come from the secondary emissions, and the corresponding
fragmentation functions are larger at low $z$. It is expected
that the $u\to K^+$ channel has a smaller probability than the $u\to\pi^+$
one does, because a kaon is more massive. The same explanation applies
to the comparison of the $u\to K^-$ ($s\to K^+$) and $u\to\pi^-$
($s\to \pi^{\pm}$) channels.

\section{GLUON FRAGMENTATION FUNCTIONS}

Due to the absence of the gluonic degrees of freedom at the Lagrangian
level in the NJL model, a gluon fragmentation function cannot be computed
directly. As stated in the Introduction, we propose to derive this
fragmentation function by treating a gluon as a pair of color lines formed
by fictitious quark and anti-quark ($q\bar q$) in a color-octet state.
The idea originates form the color dipole model developed by Gustafson and
Andersson in 80's \cite{ref35,ref36,ref37}. The large $N_c$ limit
is assumed in this model, under which parton emissions are turned
into emissions of color dipoles composed of quarks and anti-quarks in,
for instance, a shower process. The color dipole model has been also
extended to handle onium-onium scattering at high energy \cite{ref14,ref38}, for
which a high energy onium state, consisting of numerous $q\bar{q}$ pairs
and soft gluons, is regarded as a collection of color dipoles in the
large $N_c$ limit. The result has been compared with that from the
formalism with Balitskii-Fadin-Kuraev-Lipatov pomerons \cite{ref38}.

\begin{figure*}
 \centering
\includegraphics[scale=0.50]{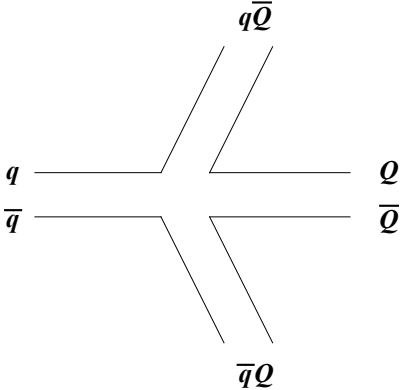}
\caption{\label{fig7}
Gluon elementary fragmentation function in the Lund model.}
\end{figure*}

A requirement is that the fictitious $q\bar q$ pair remains flavor-singlet
after meson emissions, which can be achieved by the simultaneous emissions of
$m=q\bar Q$ and $\bar m=Q\bar q$ as illustrated in Fig.~\ref{fig7}. That is, if
the $u$ quark of the $u\bar{u}$ pair fragments a $\pi^+$ meson, the $\bar{u}$
quark of the pair must fragment a $\pi^-$ meson. The $d\bar{d}$ pair after the
$\pi^+$ and $\pi^-$ emissions remains in the flavor-singlet state, and then repeats
meson emissions. Applying Fig.~\ref{fig7} to generate the jet process, we write the
resultant gluon fragmentation functions $D_g^{Lm}(z)$ as a combination of the
fragmentation functions $D_q^m(z)$ from the quark and $D_{\bar q}^m(z)$ from the
anti-quark,
\begin{equation}
\begin{aligned}
D_g^{Lm}(z)&=\sum_{q} \frac{1}{3} \int_0^1 P_{g\to q\bar{q}}(x)\left[D_q^m(\frac{z}{x})\frac{1}{x}
+D_{\bar{q}}^m(\frac{z}{1-x})\frac{1}{1-x}\right]dx,
\end{aligned}\label{dglm}
\end{equation}
for $z/x \leq 1$ in $D_q^m(z/x)$ and $z/(1-x) \leq 1$ in $D_{\bar{q}}^m(z/(1-x))$.
The gluon momentum is distributed between the quark $q$ with the
momentum fraction $x$ and the anti-quark $\bar{q}$ with $1-x$ according to the
normalized splitting function $P_{g\rightarrow q\bar{q}}$.
The average over the three fictitious quark flavors $q=u$, $d$, and $s$ has
been made explicit. Because $D_q^m$ is defined for an initial quark $q$ with
100\% momentum to fragment mesons, its argument should be rescaled, leading to
$D_q^{m}(z/x)/x$ and $D_{\bar{q}}^m(z/(1-x))/(1-x)$ in Eq.~(\ref{dglm}).
This simplest version of our proposal is consistent with the formulation of the
gluon fragmentation functions in the Lund model \cite{ref14}.

\begin{figure*}
 \centering
\includegraphics[scale=0.70]{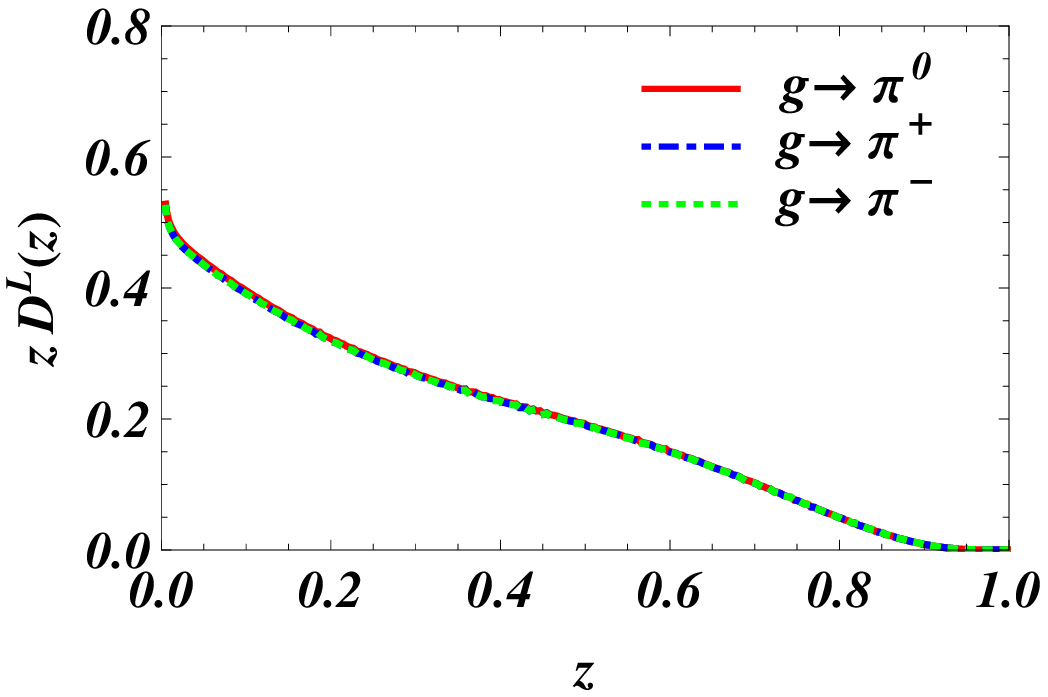}
\includegraphics[scale=0.70]{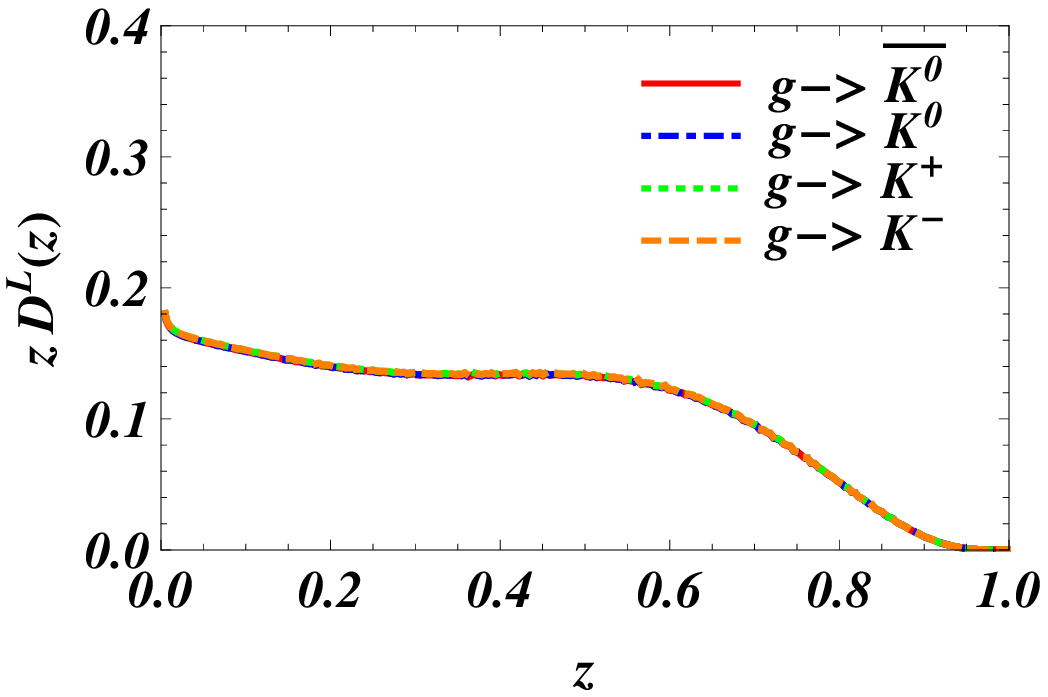}

\hspace{1.0cm}(a)\hspace{7.0cm}(b)
\caption{\label{fig8}
$z$ dependence of (a) $zD_g^{L\pi}(z)$ and (b) $zD_g^{LK}(z)$ at a model scale.}
\end{figure*}

The choice of the normalized splitting function $P_{g\rightarrow q\bar{q}}$ is
arbitrary. Fortunately, we have confirmed that our results are insensitive to
the choices of $P_{g\rightarrow q\bar{q}}$, especially after QCD evolution
effects are taken into account. Therefore, we simply assume that it
is proportional to the Dokshitzer-Gribov-Lipatov-Altarelli-Parisi (DGLAP)
kernel \cite{ref39}
\begin{equation}
\begin{aligned}
P_{g\rightarrow q\bar{q}}(x)=\frac{1}{2}(1-2x+2x^2),
\end{aligned}
\end{equation}
for $0< x< 1$. Our gluon fragmentation functions are also insensitive
to the variation of the fictitious quark masses in the involved
$d_q^m(z)$ and $d_{\bar{q}}^m$, which are then set to zero for
convenience. The values of $g_{\pi qQ}$ and $g_{KqQ}$ are the same as in the
previous section. The $z$ dependence of $zD_g^{Lm}(z)$ for a gluon fragmenting into
pions and kaons at a model scale are presented in Fig.~\ref{fig8}. The
features that the probabilities for a gluon to fragment into mesons of
different charges are identical, and that the gluon fragmentation
functions for kaons are smaller than for pions are expected. We explain
why all the gluon fragmentation functions decrease with $z$
by taking the fragmentation into the $\pi^+$ meson as an example:
the major contributions of $D_{u,\bar d}^{\pi^+}$ arise from the high $z$ region,
which is suppressed by the phase space $x\ge z$ in Eq.~(\ref{dglm}), and the
contributions of the other quark fragmentation functions are small
in the high $z$ region. It should be pointed out that $zD_g^{Lm}(z)$ vanishes
as $z\to 0$ actually, though it is hard to see this fact in Fig.~\ref{fig8}.

\begin{figure*}
 \centering
\includegraphics[scale=0.60]{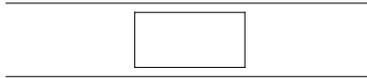}
\caption{\label{fig12}
Color lines for quark annihilation.}
\end{figure*}

A refined version of our proposal starts with the construction of
the elementary gluon fragmentation function $d_g^m(z)$, which
describes the probability of a gluon to emit a specific meson $m$ with
a momentum fraction $z$ in one step, from the elementary quark fragmentation
functions $d_q^m(z)$:
\begin{equation}
\begin{aligned}
d_{g}^m(z)&=\sum_{q} \frac{1}{3} \int_0^1 P_{g\to q\bar{q}}(x)
\left[d_q^m(\frac{z}{x})\frac{1}{x}+d_{\bar{q}}^m(\frac{z}{1-x})\frac{1}{1-x}\right]dx,
\end{aligned}\label{rdgm}
\end{equation}
for $z/x \leq 1$ in $d_q^m(z/x)$ and $z/(1-x) \leq 1$ in $d_{\bar{q}}^m(z/(1-x))$.
The essential difference of the above construction from the Lund model is that each
meson emission by a gluon has no correlation with the previous one:
once the quark annihilation mechanism depicted in Fig.~\ref{fig12} is
combined with Fig.~\ref{fig7}, the quark flavor at each emission is
arbitrary (it could be $u$, $d$, or $s$). Namely, the specific flavor
of the fictitious $q\bar q$ pair is irrelevant, and the color lines mainly provide
color sources of meson emissions.

\begin{figure*}
 \centering
\includegraphics[scale=0.55]{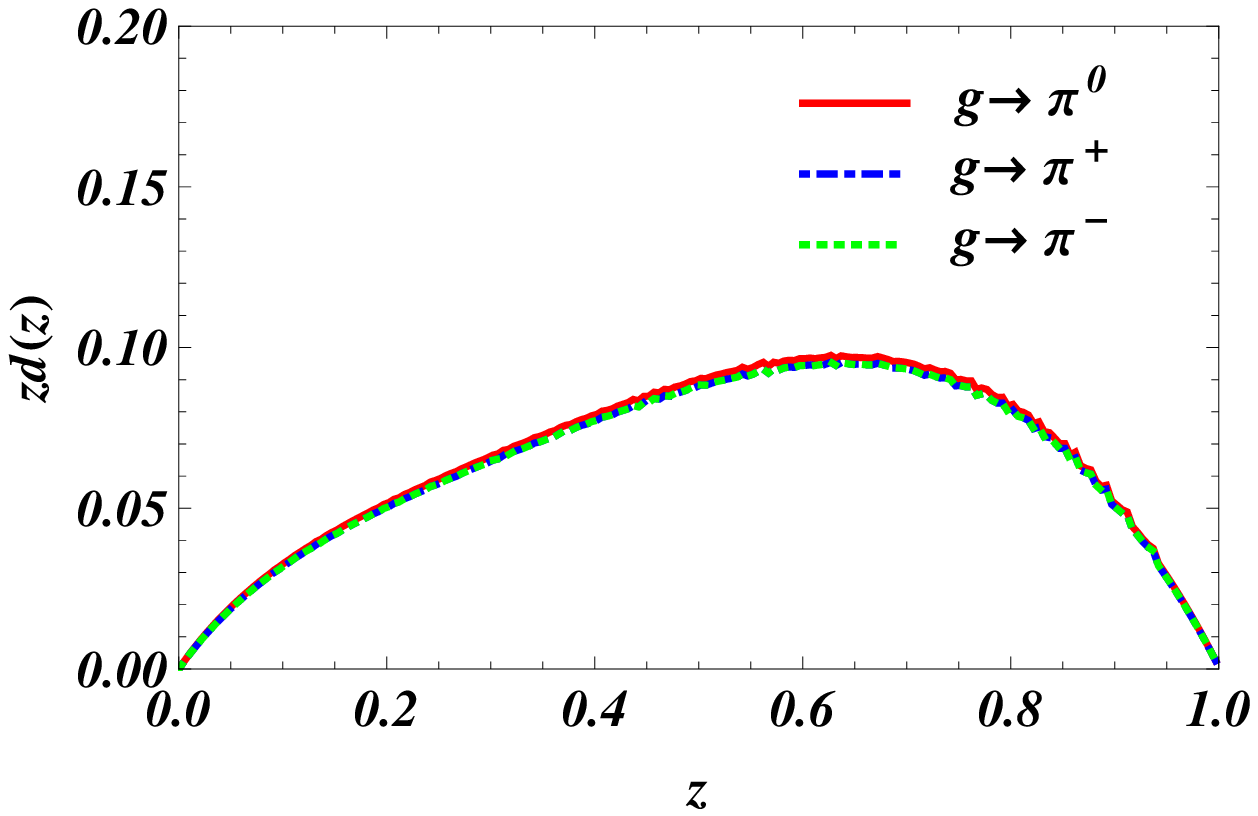}
\includegraphics[scale=0.55]{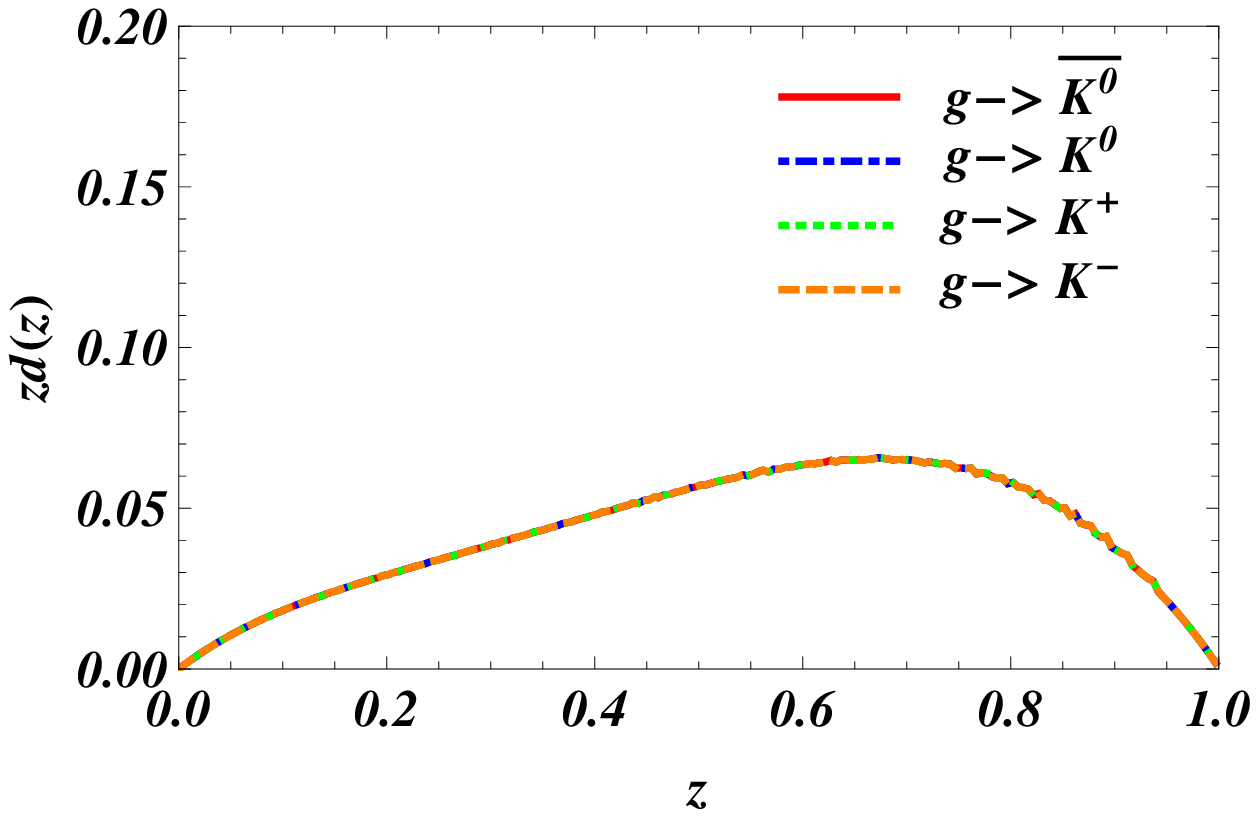}

\hspace{1.0cm}(a)\hspace{7.0cm}(b)
\caption{\label{fig11}
$z$ dependence of (a) $zd_g^{\pi}(z)$ and (b) $zd_g^{K}(z)$.}
\end{figure*}

The $z$ dependence of $zd_{g}^m(z)$ for the one-step fragmentation of a gluon
into pions and kaons is displayed in Fig.~\ref{fig11}. Similarly, the probabilities
for fragmenting into mesons of different charges by a gluon are identical as expected,
and the elementary fragmentation functions for kaons are smaller than for pions.
The comparison between the behaviors of the gluon and quark elementary fragmentation
functions is also similar to the comparison between the gluon and quark fragmentation
functions in the Lund model. In the present case the quark elementary fragmentation
functions $zd_{q}^m(z)$ vanish quickly at low $z$ as indicated in Fig.~\ref{fig4},
such that $zd_{g}^m(z)$ also vanish at low $z$, and have peaks at high $z$.

\begin{figure*}
 \centering
\includegraphics[scale=0.70]{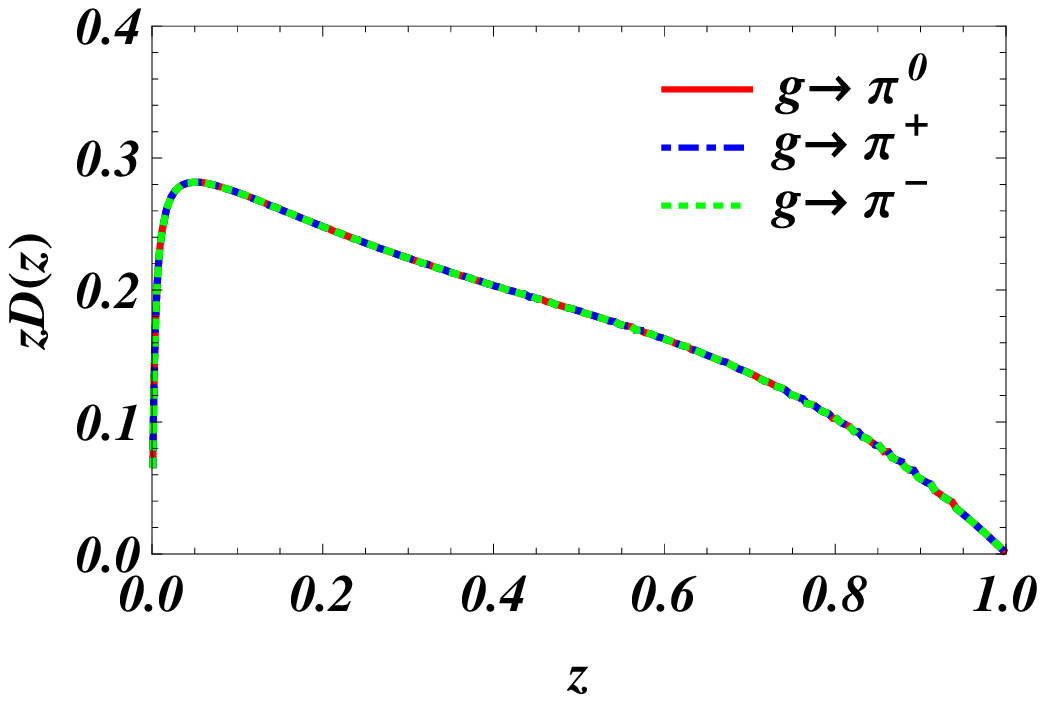}
\includegraphics[scale=0.70]{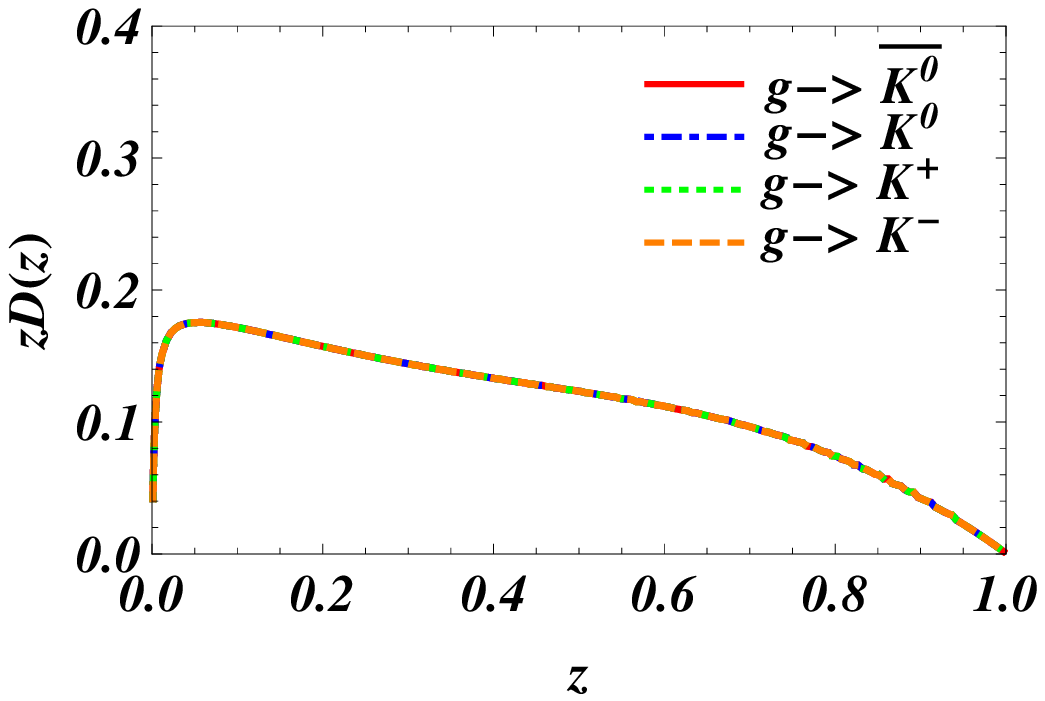}

\hspace{1.0cm}(a)\hspace{7.0cm}(b)
\caption{\label{fig13}
$z$ dependence of (a) $zD_g^{\pi}(z)$ and (b) $zD_g^{K}(z)$ at a model scale.}
\end{figure*}

Since gluons before and after meson emissions are represented by the
pair of color lines without referring to specific quark flavors in the above
construct (the memory of specific quark flavors has been washed out by the
introduction of the quark annihilation mechanism), the gluon fragmentation
function $D_g^m(z)$ satisfies the integral equation
\begin{equation}
\begin{aligned}
D_{g}^m(z)=&\hat{d}_{g}^m(z)+\sum_{m'}{\int_z^1\frac{dy}{y}\hat{d}_{g}^{m'}(1-y)
D_{g}^m(\frac{z}{y})}.
\label{njli}
\end{aligned}
\end{equation}
Note that $d_{g}^m(z)$ has been normalized into $\hat{d}_{g}^m(z)$ in order to
have a probability meaning, and $\hat{d}_{g}^{m'}(1-y)$ is interpreted as the
probability $\hat{d}_{g}^{g}(y)$. The solutions of $zD_g^m(z)$
to Eq.~(\ref{njli}) at a model scale are collected in Fig.~\ref{fig13}. Compared
to the results from the Lund model in Fig.~\ref{fig8}, the most significant
difference appears in the region of $z<0.2$, where $zD_g^m(z)$
grow more slowly as $z$ decreases, and descend to zero as $z\to 0$ more
quickly than $zD_g^{Lm}(z)$ do. This difference is attributed to the flavor
blindness of the color lines, which renders meson emissions easier and shifts
the peaks of the gluon fragmentation functions to a bit higher $z$.

\begin{figure*}
 \centering
\includegraphics[scale=0.55]{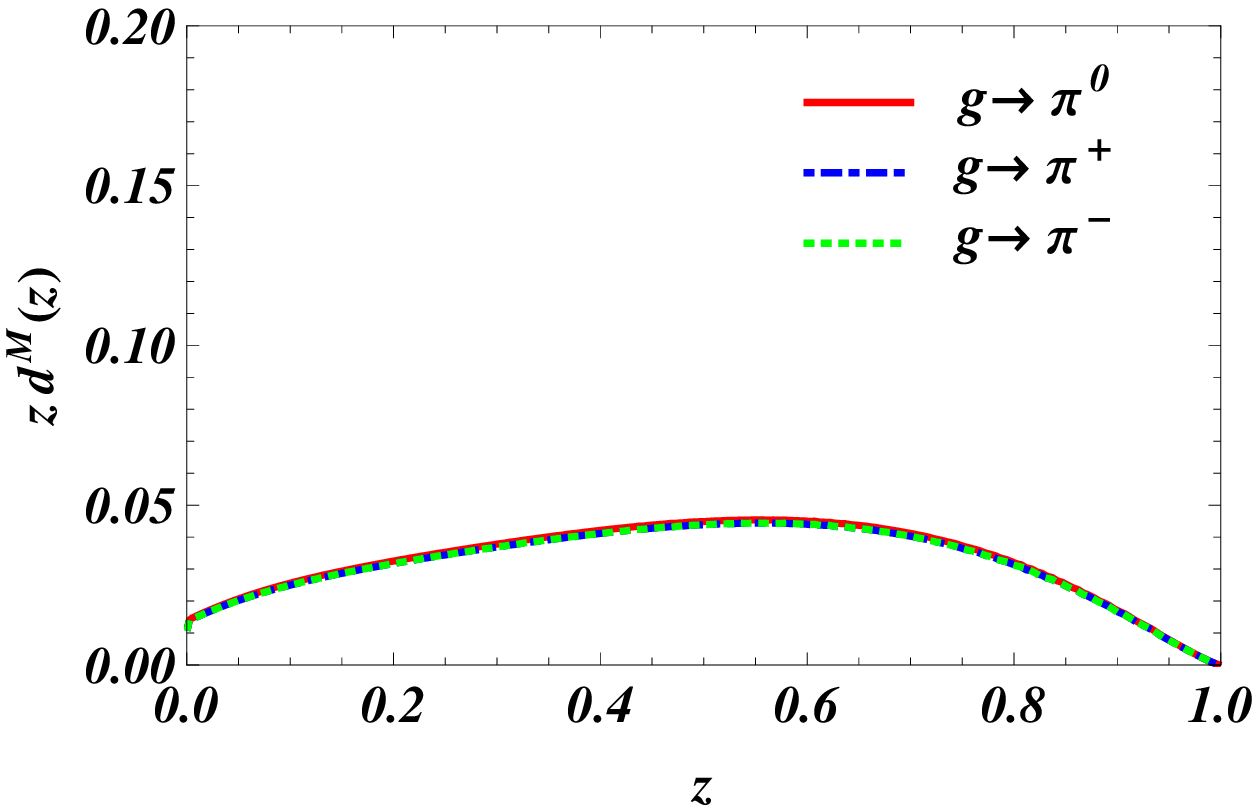}
\includegraphics[scale=0.55]{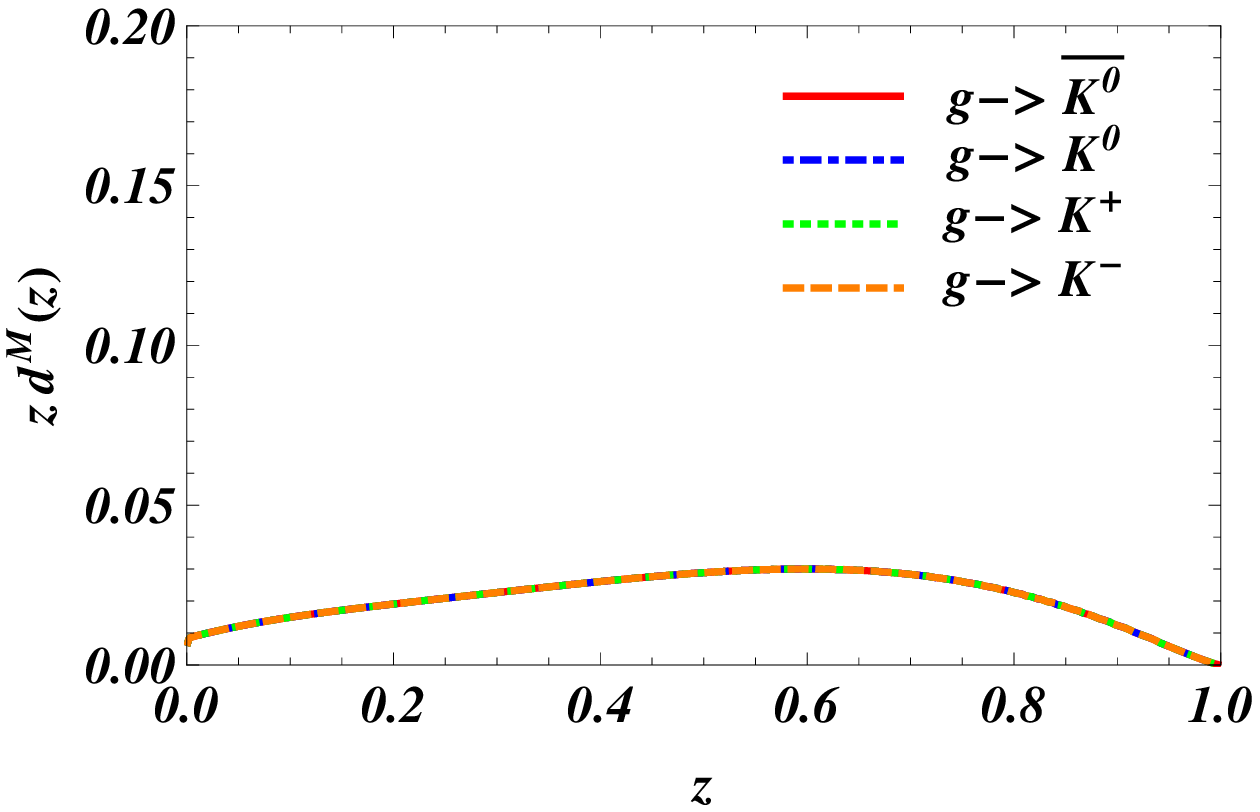}

\hspace{1.0cm}(a)\hspace{7.0cm}(b)
\caption{\label{fig14}
$z$ dependence of (a) $zd_{g}^{M\pi}(z)$ and (b) $zd_{g}^{MK}(z)$.}
\end{figure*}

\begin{figure*}
 \centering
\includegraphics[scale=0.55]{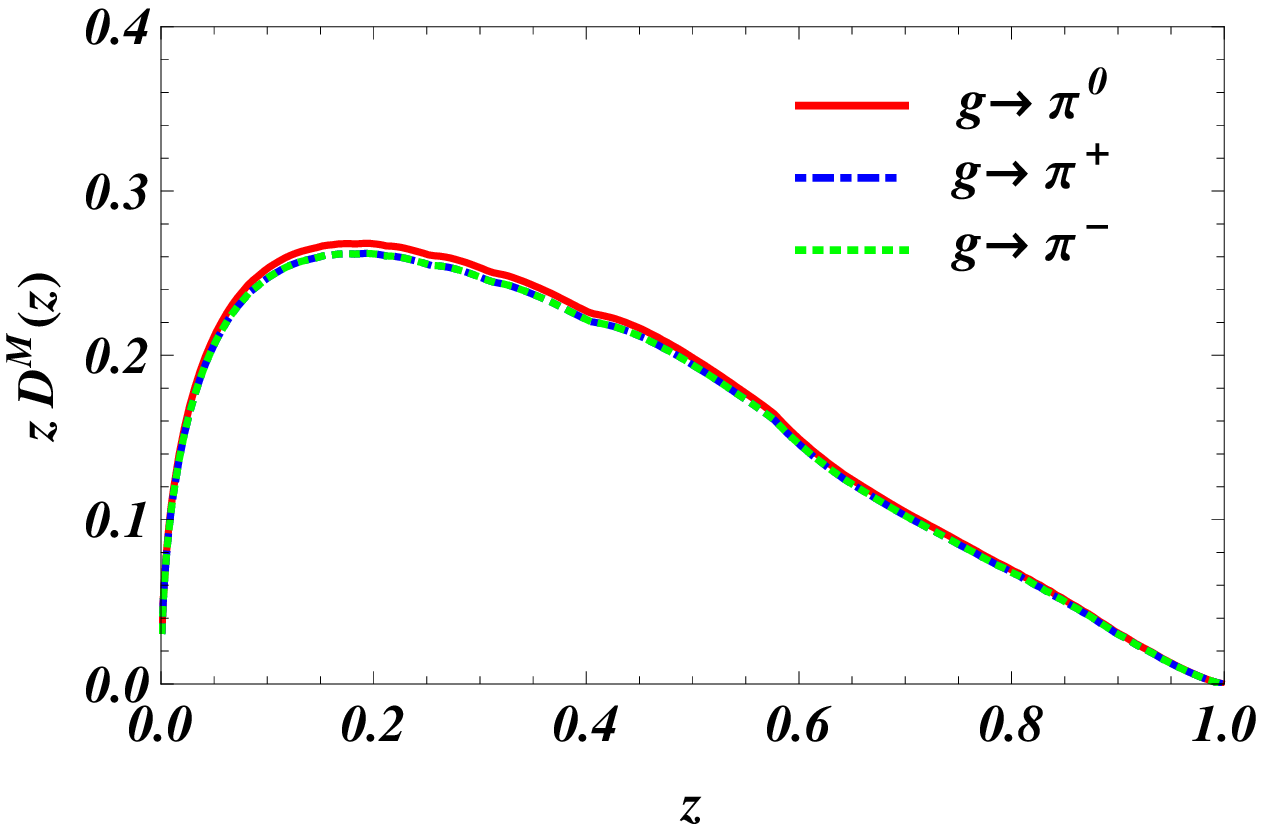}
\includegraphics[scale=0.55]{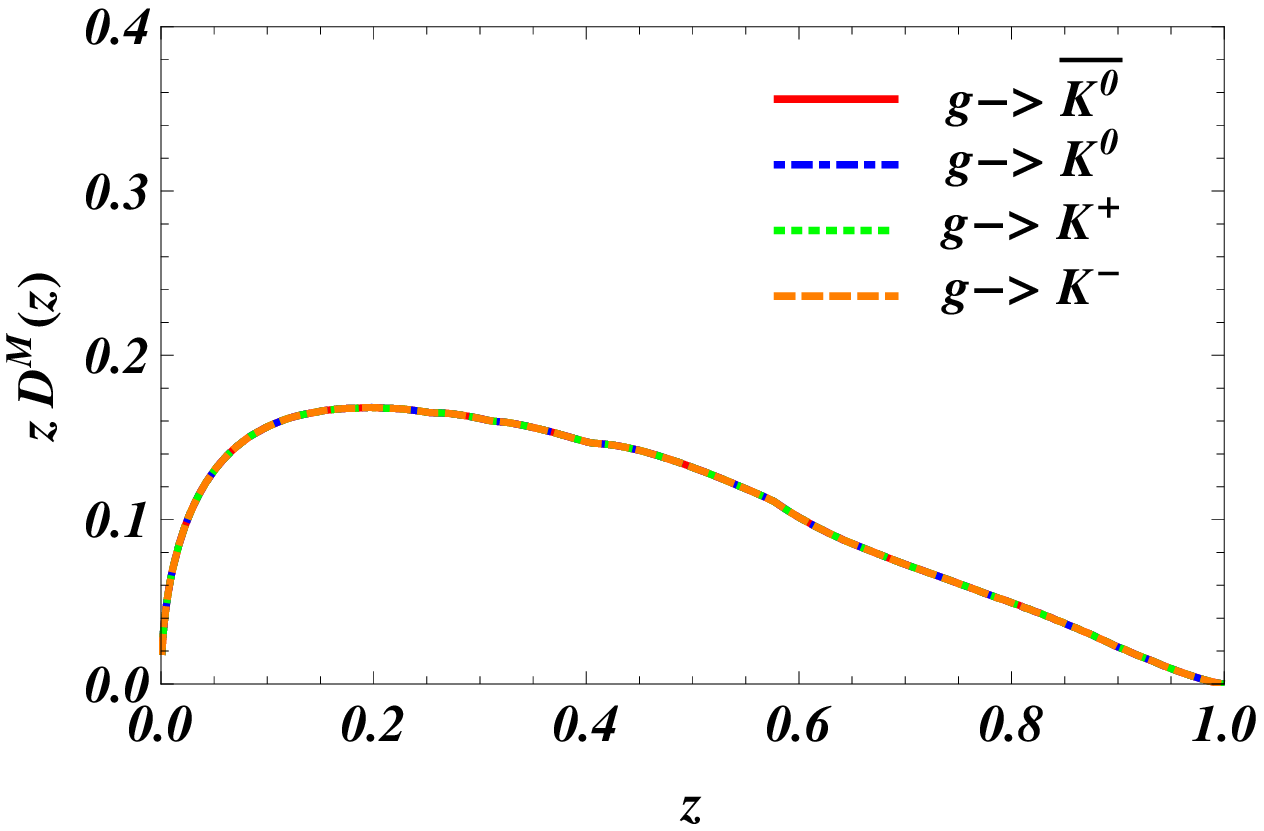}

\hspace{1.0cm}(a)\hspace{7.0cm}(b)
\caption{\label{fig15}
$z$ dependence of (a) $zD_{g}^{M\pi}(z)$ and (b) $zD_{g}^{MK}(z)$
at a model scale.}
\end{figure*}

The gluonic dynamics is more complicated than discussed above definitely.
For instance, the fictitious quark pair can split into two or more fictitious
quark pairs at any stage of meson emissions. To test the impact of this
multi-dipole mechanism, we consider a more complicated elementary gluon
fragmentation function $d_{g}^{Mm}(z)$ in terms of $d_g^m(z)$ in Eq.~(\ref{rdgm}):
\begin{equation}
\begin{aligned}
d_{g}^{Mm}(z)&=\int_0^1 P_{g\to gg}(x)
\left[d_{g}^m(\frac{z}{x})\frac{1}{x}+d_{g}^m(\frac{z}{1-x})\frac{1}{1-x}\right]dx,
\end{aligned}
\end{equation}
for $z/x \leq 1$ in $d_{g}^m(z/x)$ and $z(1-x) \leq 1$ in $d_{g}^m(z/(1-x))$.
As a test, the splitting function $P_{g\to gg}(x)$ is simply chosen to be proportional
to the DGLAP kernel
\begin{equation}
\begin{aligned}
P_{g\rightarrow gg}(x)=6\left[\frac{1-x}{x}+x(1-x)+\frac{x}{1-x}\right],
\end{aligned}
\end{equation}
for $0<x<1$. The $z$ dependence of $zd_{g}^{Mm}(z)$ in Fig.~\ref{fig14}
is basically similar to that of $zd_{g}^{m}(z)$ in Fig.~\ref{fig11}, but more flat.
The flatness of $zd_{g}^{Mm}(z)$ makes the curves of $zD_{g}^{Mm}(z)$ in
Fig.~\ref{fig15} more smooth in the low $z$ region, compared to the curves
of $zD_{g}^m(z)$ in Fig.~\ref{fig13}.

A remark is in order. To keep the fictitious quark pair in the flavor-singlet state,
the anti-quark must emit a $\pi^-$ meson, as the quark emits a $\pi^+$ meson. One
may wonder about the channel with only one of the quarks emitting a $\pi^0$ meson, which
does not defy the flavor-singlet requirement. This $\pi^0$ emission
seems to enhance the neutral pion production over the charged ones. In fact, this
$g\to g\pi^0$ process should be regarded as the splitting of the quark pair
into two quark pairs, whose contribution has been taken into account in the more
complicated gluon fragmentation function $D_{g}^{Mm}(z)$. Hence, the probabilities
for a gluon to fragment into charged and neutral pions will be always equal in
our approach.

\section{COMPARISON WITH DATA}

We have established the gluon fragmentation functions in the NJL model at a
model scale, at which the momentum sum rule of those from the scheme
consistent with the Lund model gives $1.0072$,
the sum rule of those including the quark annihilation mechanism gives $0.9612$,
and the sum rule of those including the multi-dipole contribution gives $0.9045$.
All of them are close to unity, implying that our numerical analysis is reliable.
The plots presented in the previous section show that the gluon
fragmentation functions for the charged and neutral pions are the same and
those for the four types of kaons are the same too, so we will investigate only the
cases of $\pi^+$ and $K^+$ productions here.

We examine the behaviors of the quark and gluon fragmentation functions in different
schemes under the LO and NLO QCD evolutions from the model scale $Q_0^2=0.15$
GeV$^2$ and $Q_0^2=0.17$ GeV$^2$, respectively, to higher scales. The model scales,
being free parameters, are chosen to attempt a reasonable fit of the predicted cross
section to the SLD data at $Q^2=M_Z^2$. Note that the
model scale for the derivation of the quark fragmentation functions in the NJL model
was set to 0.2 GeV$^2$ in \cite{ref32}. For the study of the
NLO evolution effect, we adopt the code QCDNUM \cite{ref40}. Since the observations
made from the LO and NLO evolutions are similar, we present only the
results of the latter, and collect the former ones in the Appendix.
It is worth mentioning that the momentum sum rule for a fragmentation function
is indeed violated under the QCD evolution, as postulated in the Introduction,
if the gluon fragmentation function was assumed to be null at the model scale:
we get $\sum_h{\int zD_u^h(z)dz=0.6488}$ and
$\sum_h{\int zD_g^h(z)dz=0.1929}$ at $Q^2=4$ GeV$^2$ under the LO evolution in
this case. After including the gluon fragmentation functions, the above values
are improved into  $\sum_h{\int zD_u^h(z)dz=0.9623}$ and $\sum_h{\int zD_g^h(z)dz=0.9334}$.

\begin{figure*}
 \centering
\includegraphics[scale=0.7]{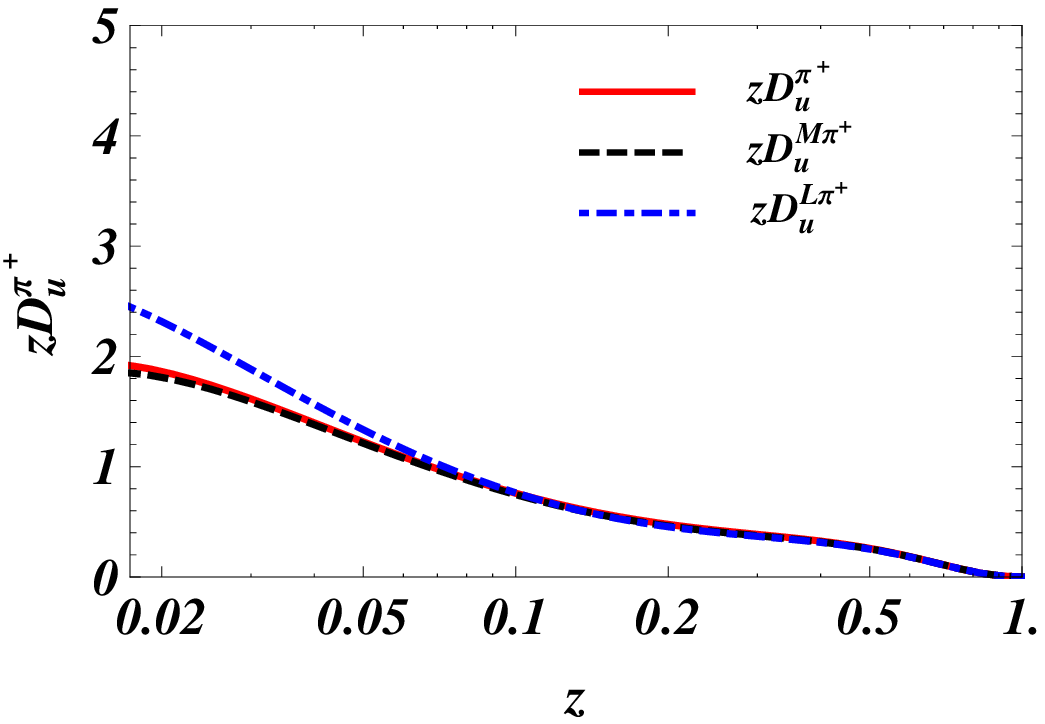}
\includegraphics[scale=0.7]{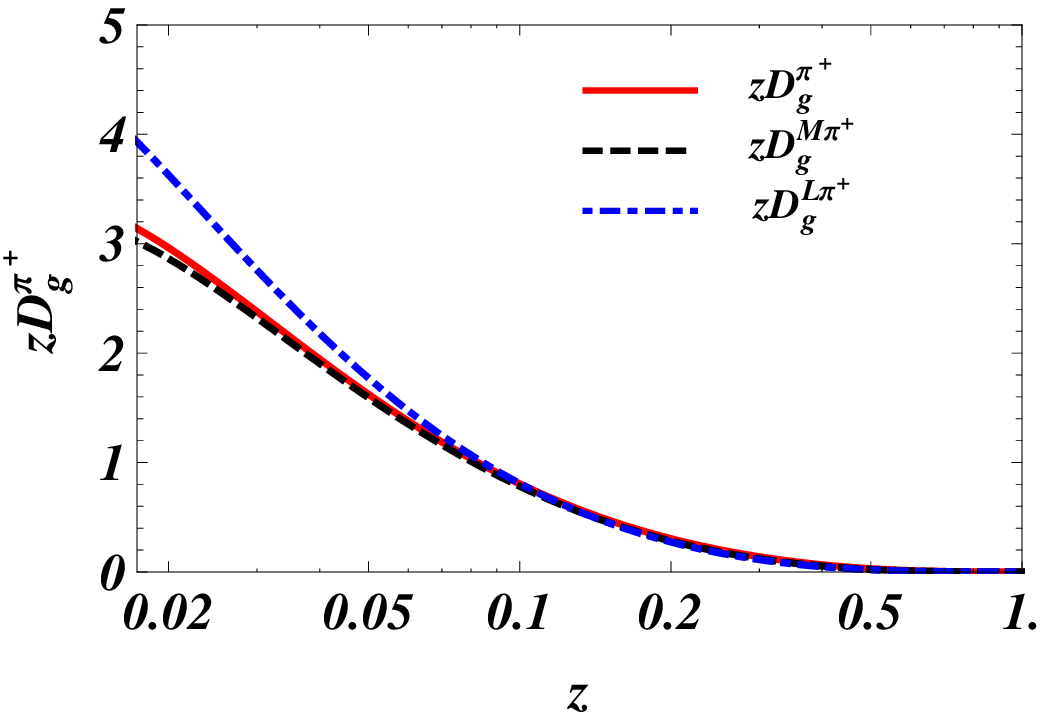}

\hspace{1.0cm}(a)\hspace{7.0cm}(b)
\includegraphics[scale=0.7]{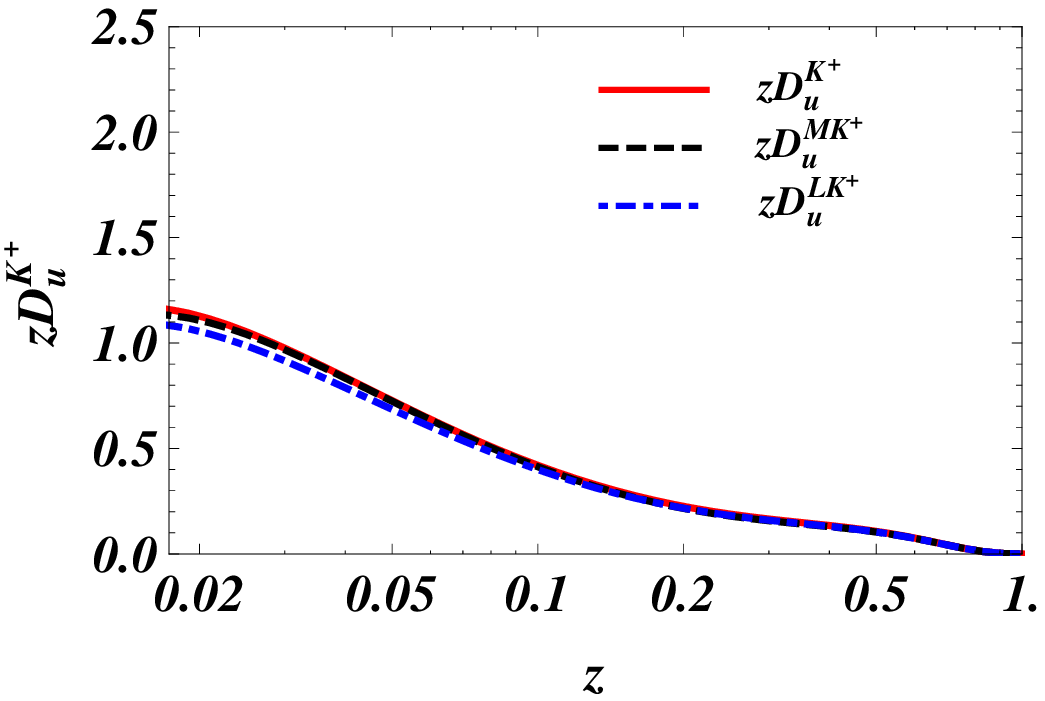}
\includegraphics[scale=0.7]{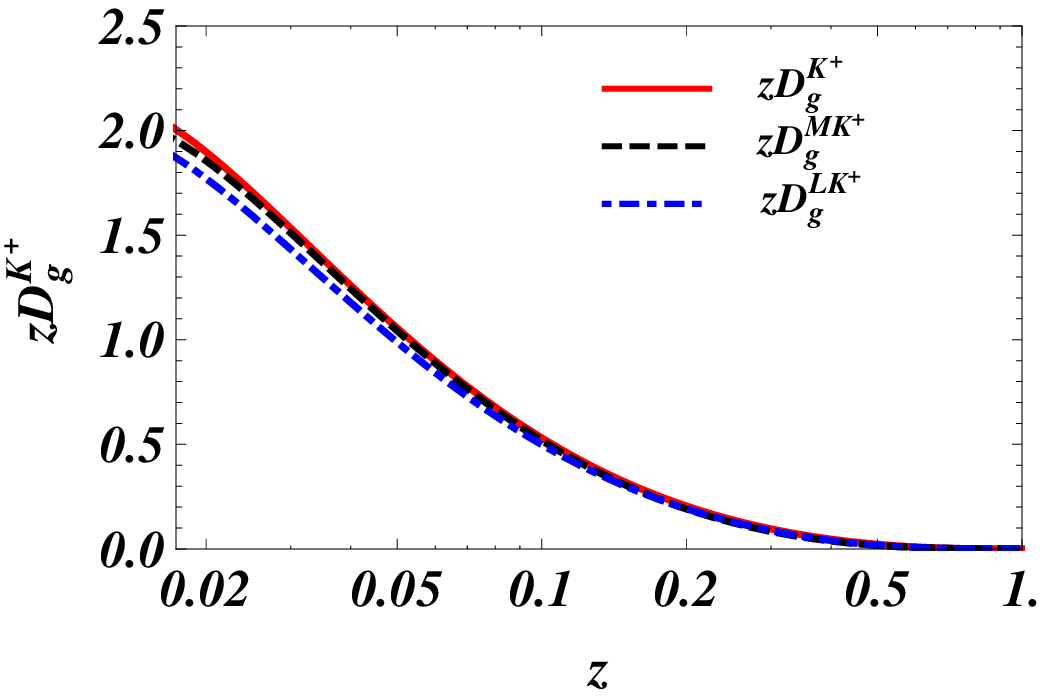}

\hspace{1.0cm}(c)\hspace{7.0cm}(d)
\caption{\label{fig21}
$z$ dependence of (a) $zD_{u}(z)$, $zD_{u}^{M}(z)$, and $zD_{u}^{L}(z)$, and
(b) $zD_{g}(z)$, $zD_{g}^{M}(z)$, and $zD_{g}^{L}(z)$ for the $\pi^+$ meson emission
at the scale $Q^2=4$ GeV$^2$ under the NLO evolution. (c) and (d) are for
the $K^+$ meson emission.}
\end{figure*}

The $u$-quark and gluon fragmentation functions from the three different schemes
at $Q^2 =4$ GeV$^2$ under the NLO QCD evolution are compared in Fig.~\ref{fig21}.
The four plots indicate that the evolution effect pushes the difference among the
three schemes of handling subtle gluonic dynamics to the region of very small
$z<0.05$. We expect that the difference of the quark and gluon fragmentation functions
will move into the region of even lower $z$, as $Q^2$ increases up to $M_Z^2$.
This explains why our results are stable with respect to the variation of model
parameters and to the choices of the splitting functions. Besides, the similarity of
the curves for $zD$ and $zD^M$ hints that the gluon branching effect may not
be crucial. Therefore, it suffices to concentrate only on the scheme with the quark
annihilation mechanism below in the scope of the present work.

We then compare our results for the $\pi^+$ emission at $Q^2=4$ GeV$^2$ with
the HKNS \cite{ref26} and DSS \cite{ref27} parameterizations, whose initial scale
was set to $Q^2 = 1$ GeV$^2$, under the NLO QCD evolution in
Fig.~\ref{fig23}, and for the $K^+$ emission in Fig.~\ref{fig25}. The
comparison at the scale $Q^2=M_Z^2$ for the $\pi^+$ and $K^+$ emissions
is made in Figs.~\ref{fig27} and \ref{fig29}, respectively. In the above
plots, the label "NJL without g" in the legend refers to the curves with the
gluon fragmentation functions set to zero at the model scale, and the label
"NJL with g" refers to the curves including the contribution of the gluon
fragmentation functions. Note that the HKNS and DSS parameterizations,
extracted from different sets of data, may differ quite a bit in some channels,
especially in the low $z$ region. Hence, the comparison just means to give a
rough idea on the behaviors of these fragmentation functions obtained in the
literature. These figures exhibit obvious difference between the curves labeled
by "NJL with g" and by "NJL without g" at $Q^2=4 GeV^2$ and $Q^2=M_Z^2$,
implying the importance of the gluon fragmentation functions. At both energy
scales, the curves for all the $\pi^+$ meson channels labeled by "NJL with g"
are closer to the HKNS or DSS ones than those labeled by "NJL without g"
in the almost entire region of $z$. For the $K^+$ meson channels at both
$Q^2=4$ GeV$^2$ and $Q^2= M_Z^2$, it is hard to tell which curves,
"NJL with g" or "NJL without g", are closer to the HKNS and DSS ones. However,
the "NJL with g" ("NJL without g") curves seem to be closer to the HKNS (DSS)
ones at $Q^2=M_Z^2$. It is a general trend that all the curves are more
distinct in the low $z$ region.

\begin{figure*}
 \centering
\includegraphics[scale=0.7]{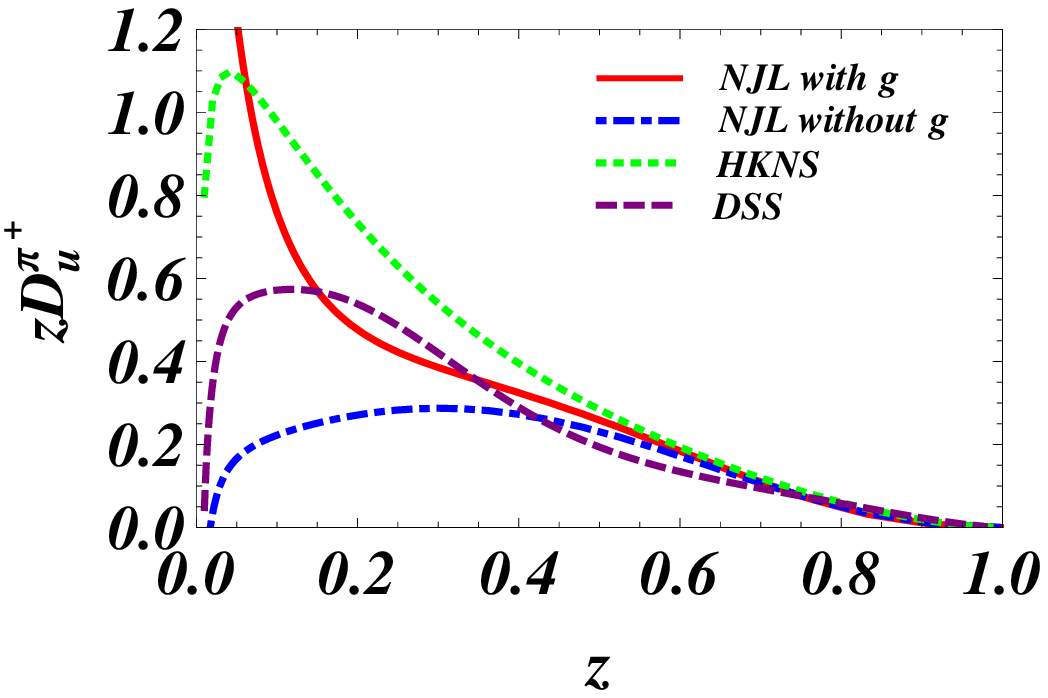}
\includegraphics[scale=0.7]{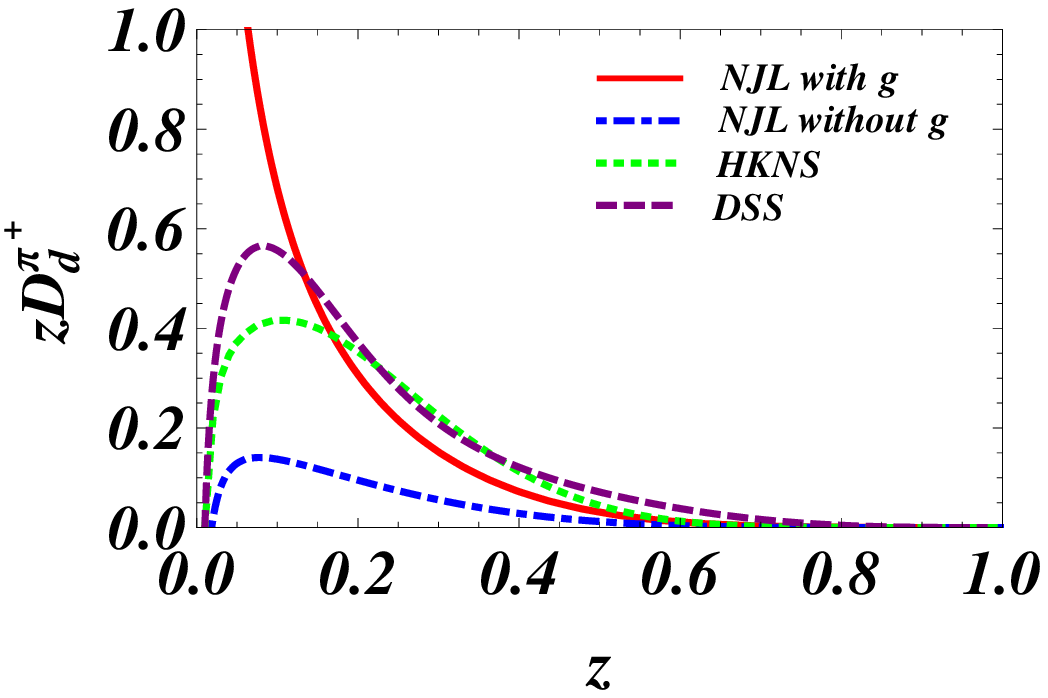}

\hspace{1.0cm}(a)\hspace{7.0cm}(b)
\includegraphics[scale=0.7]{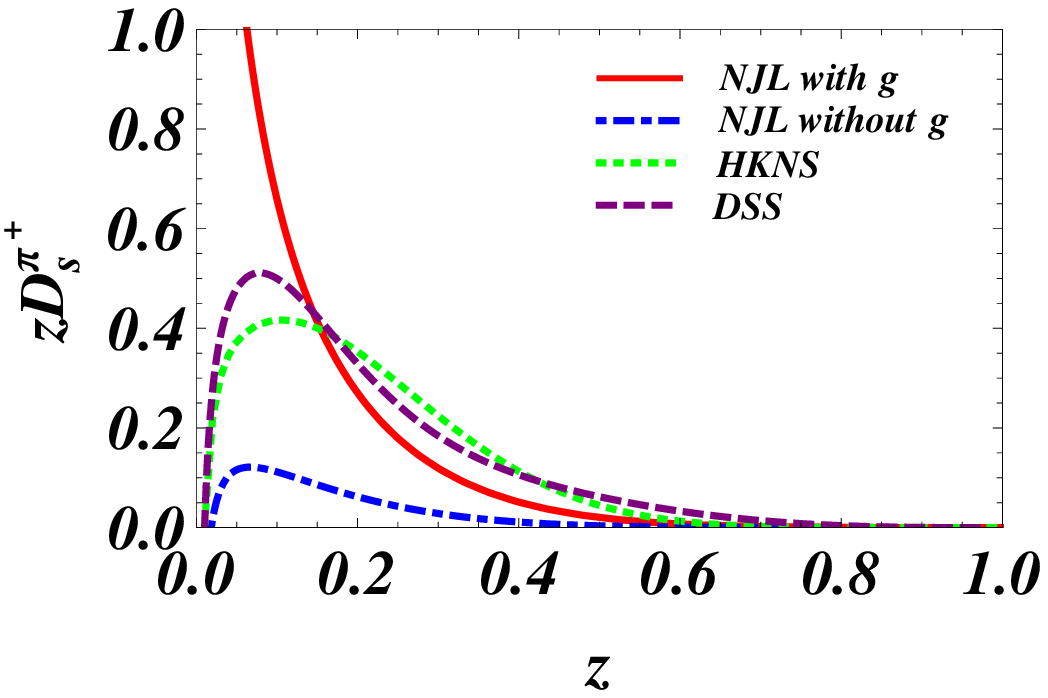}
\includegraphics[scale=0.7]{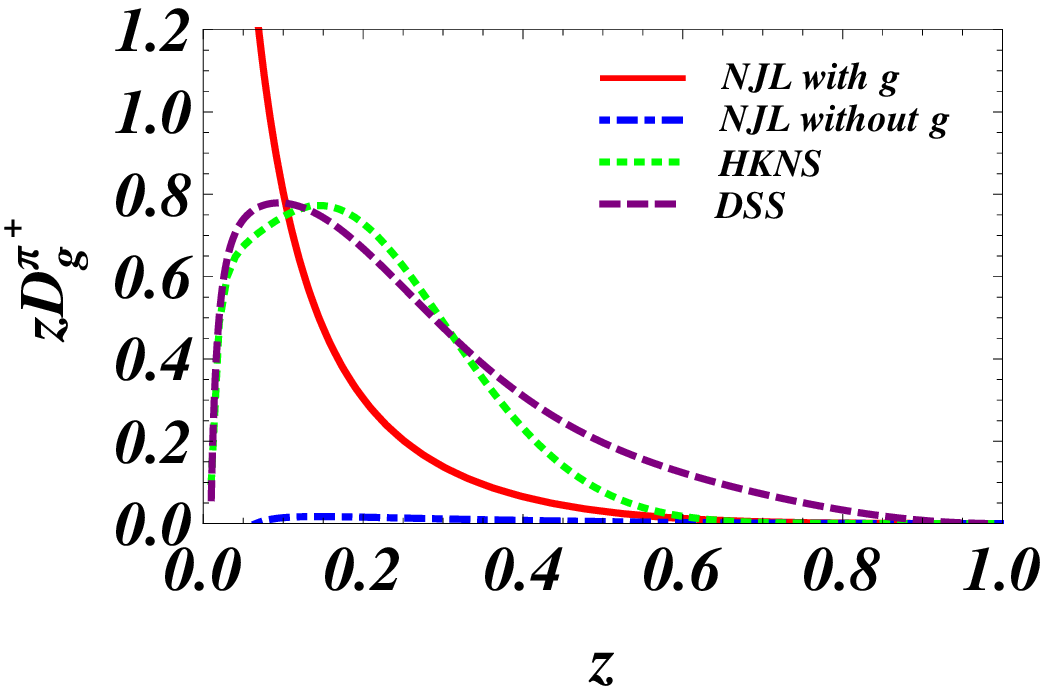}

\hspace{1.0cm}(c)\hspace{7.0cm}(d)
\caption{\label{fig31}
Comparison of (a) $zD_{u}^{\pi^+}(z)$, (b) $zD_{d}^{\pi^+}(z)$, (c) $zD_{s}^{\pi^+}(z)$,
and (d) $zD_{g}^{\pi^+}(z)$ with the HKNS and DSS parameterizations at the scale
$Q^2=4$ GeV$^2$ under the NLO evolution.}
\end{figure*}

\begin{figure*}
 \centering
\includegraphics[scale=0.7]{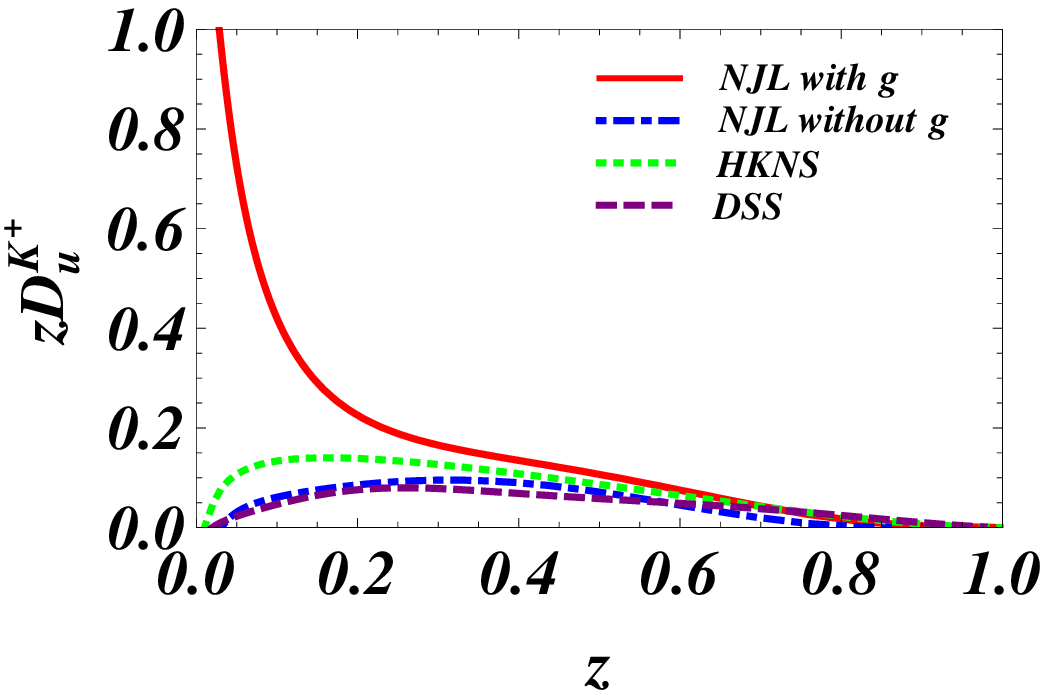}
\includegraphics[scale=0.7]{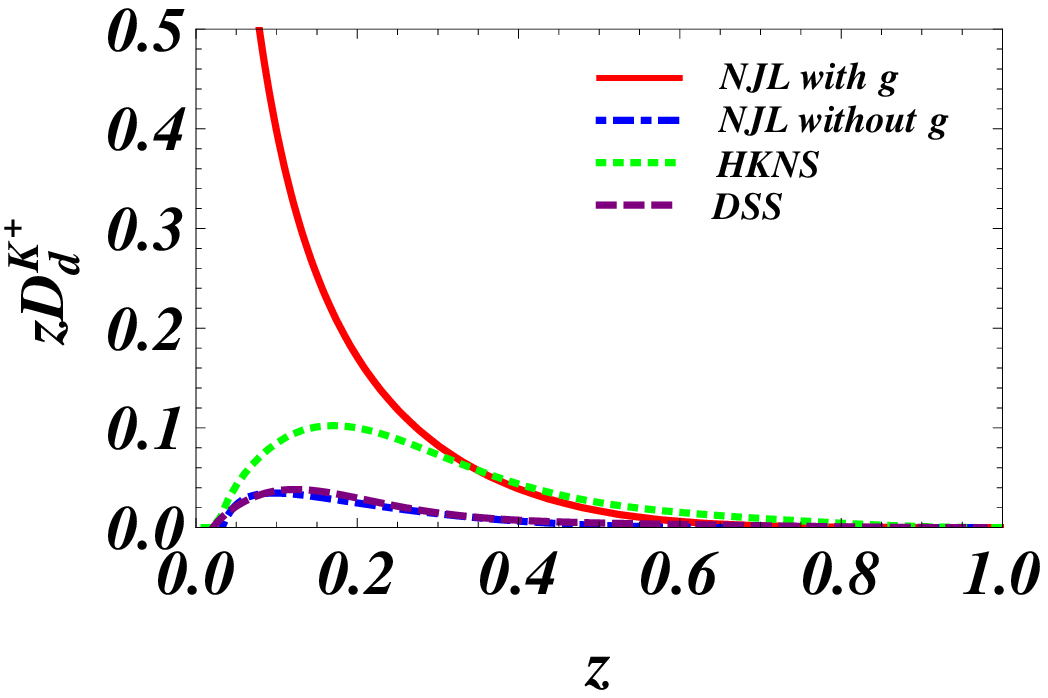}

\hspace{1.0cm}(a)\hspace{7.0cm}(b)
\includegraphics[scale=0.7]{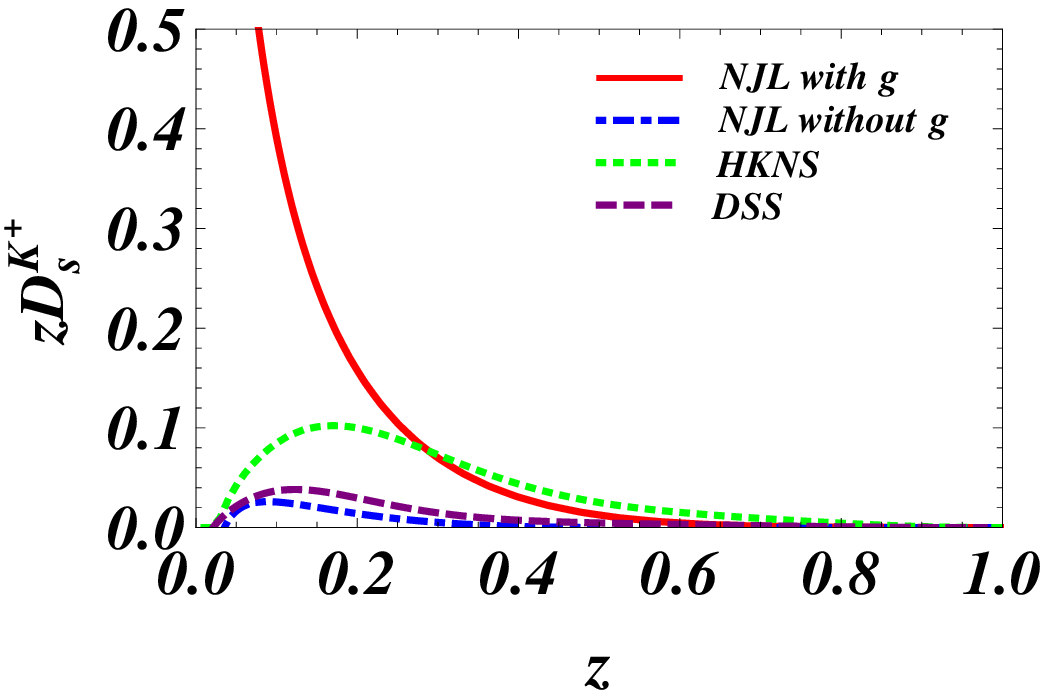}
\includegraphics[scale=0.7]{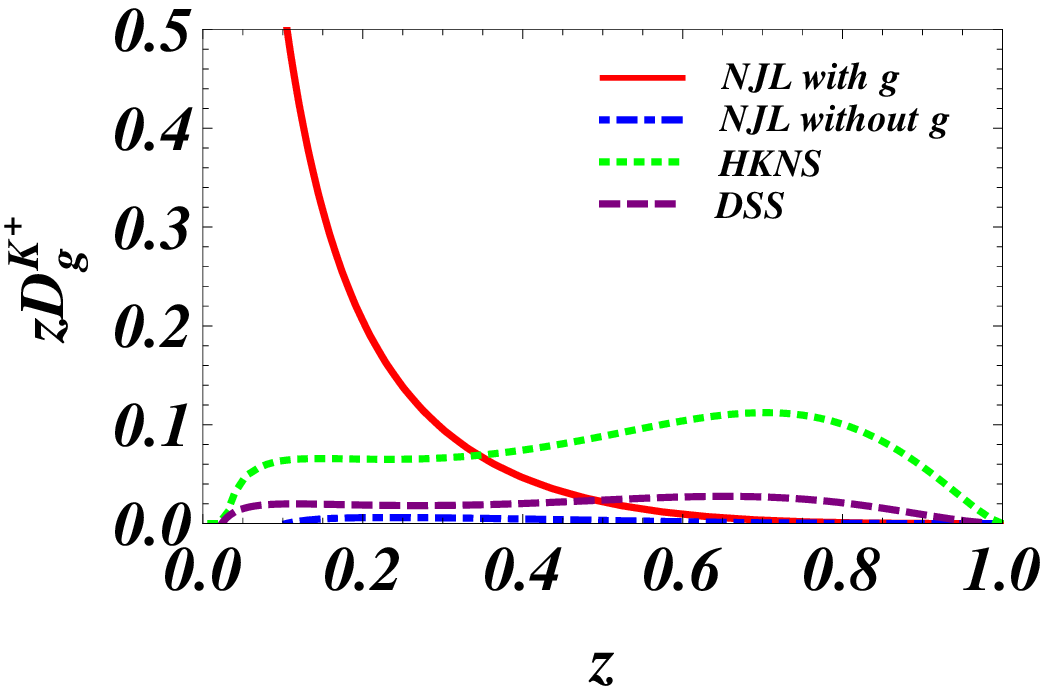}

\hspace{1.0cm}(c)\hspace{7.0cm}(d)
\caption{\label{fig33}
Same as Fig.~\ref{fig31}, but for the $K^+$ meson emission.}
\end{figure*}

\begin{figure*}
 \centering
\includegraphics[scale=0.7]{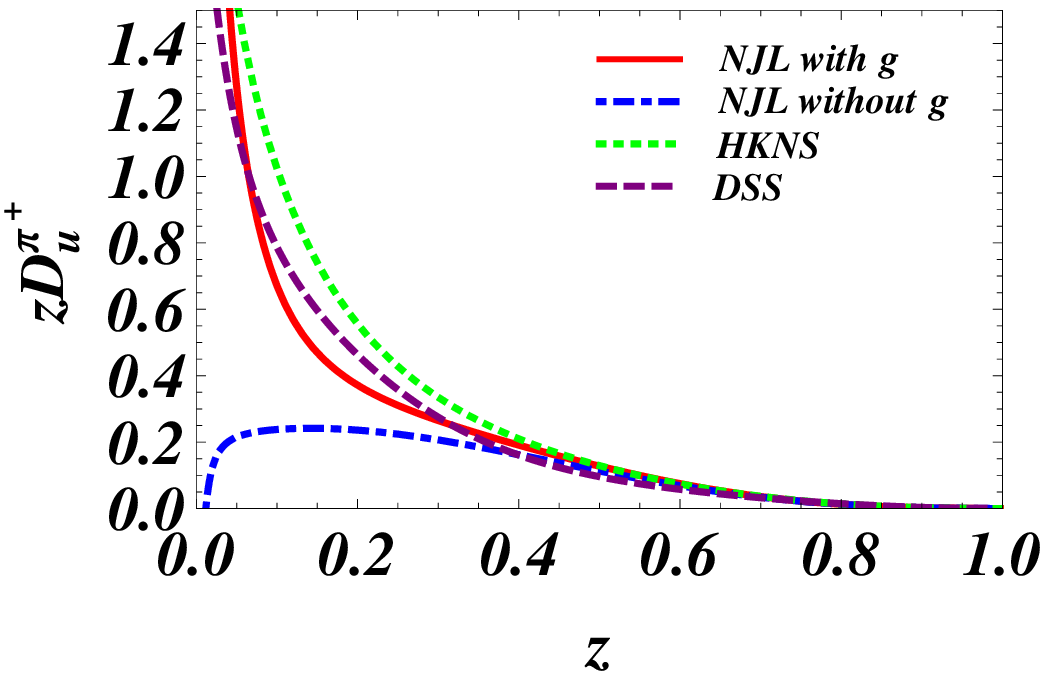}
\includegraphics[scale=0.7]{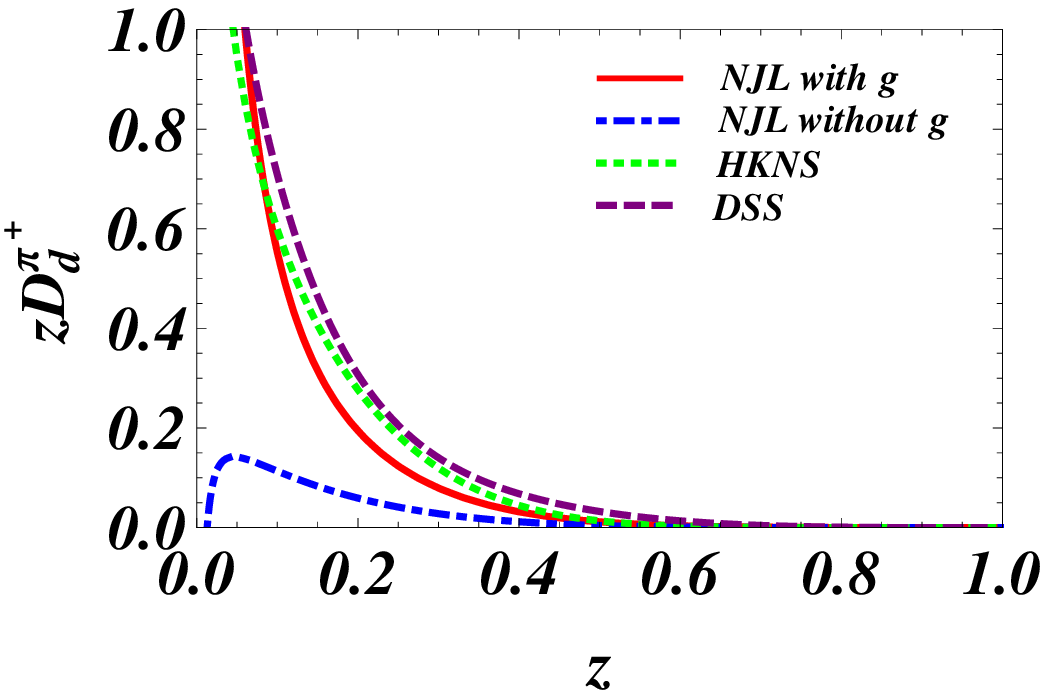}

\hspace{1.0cm}(a)\hspace{7.0cm}(b)
\includegraphics[scale=0.7]{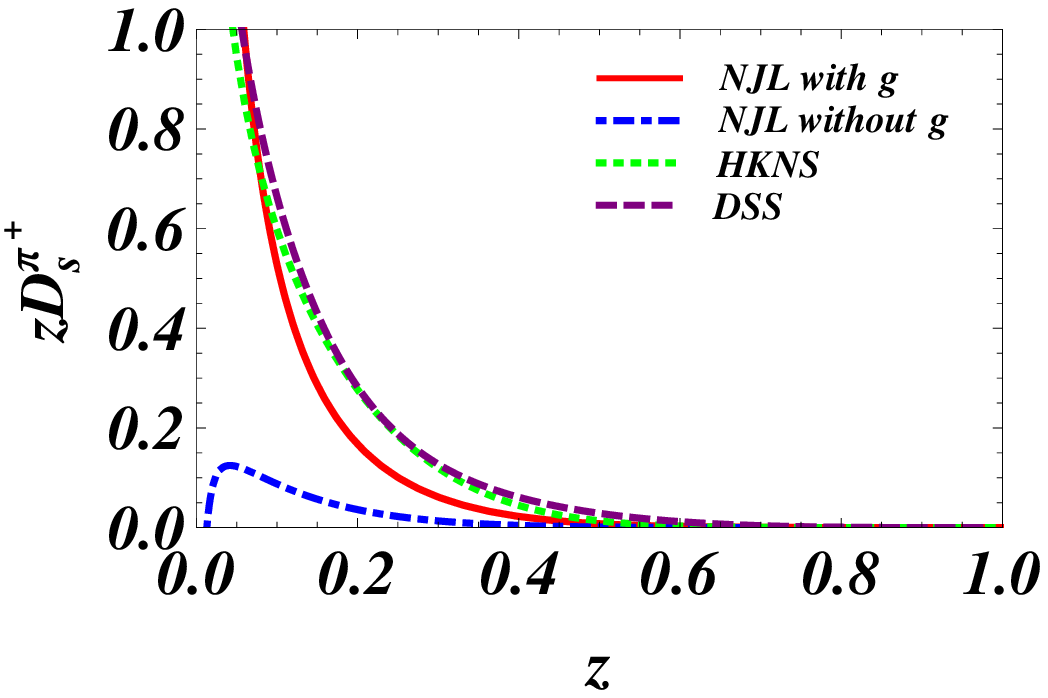}
\includegraphics[scale=0.7]{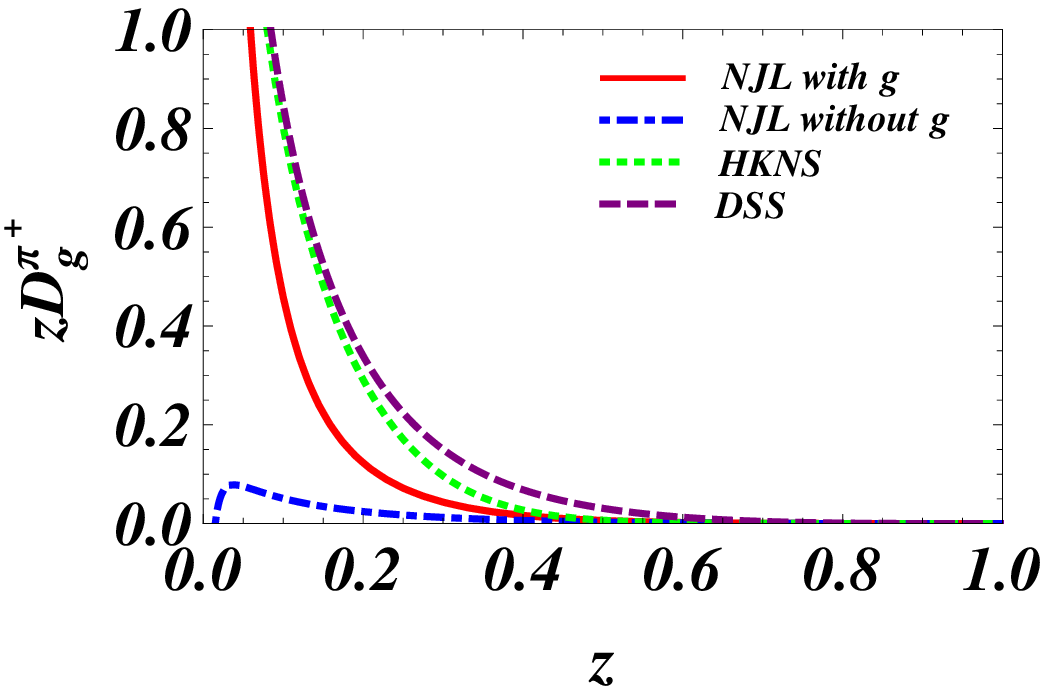}

\hspace{1.0cm}(c)\hspace{7.0cm}(d)
\caption{\label{fig35}
Comparison of (a) $zD_{u}^{\pi^+}(z)$, (b) $zD_{d}^{\pi^+}(z)$, (c) $zD_{s}^{\pi^+}(z)$,
and (d) $zD_{g}^{\pi^+}(z)$ with the HKNS and DSS parameterizations at the scale
$Q^2=M_Z^2$ under the NLO evolution.}
\end{figure*}

\begin{figure*}
 \centering
\includegraphics[scale=0.7]{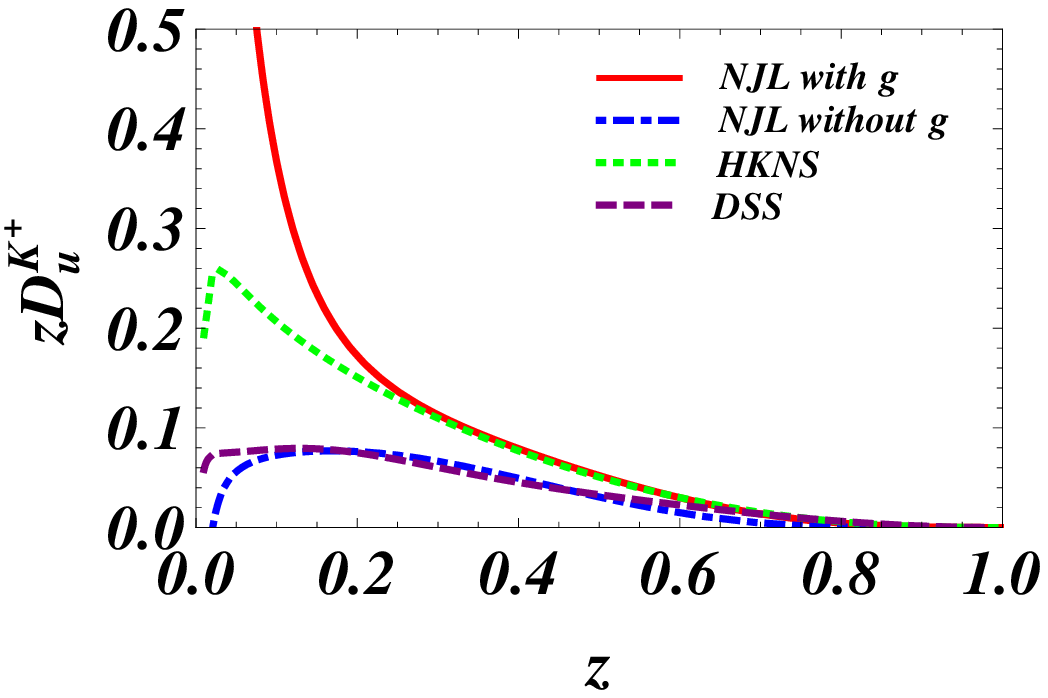}
\includegraphics[scale=0.7]{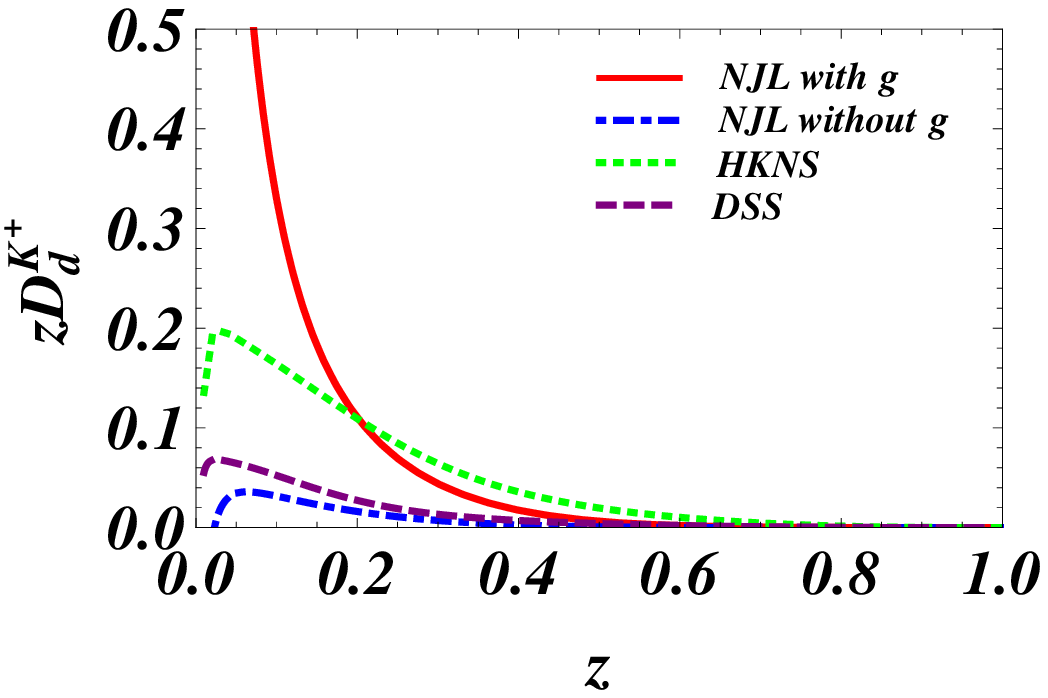}

\hspace{1.0cm}(a)\hspace{7.0cm}(b)
\includegraphics[scale=0.7]{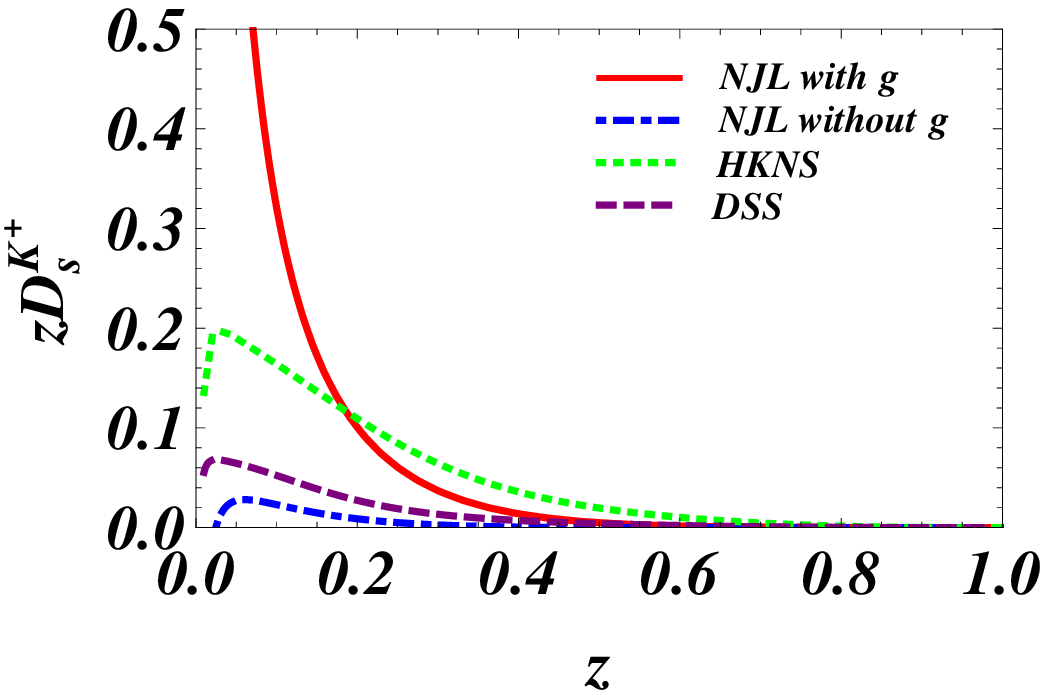}
\includegraphics[scale=0.7]{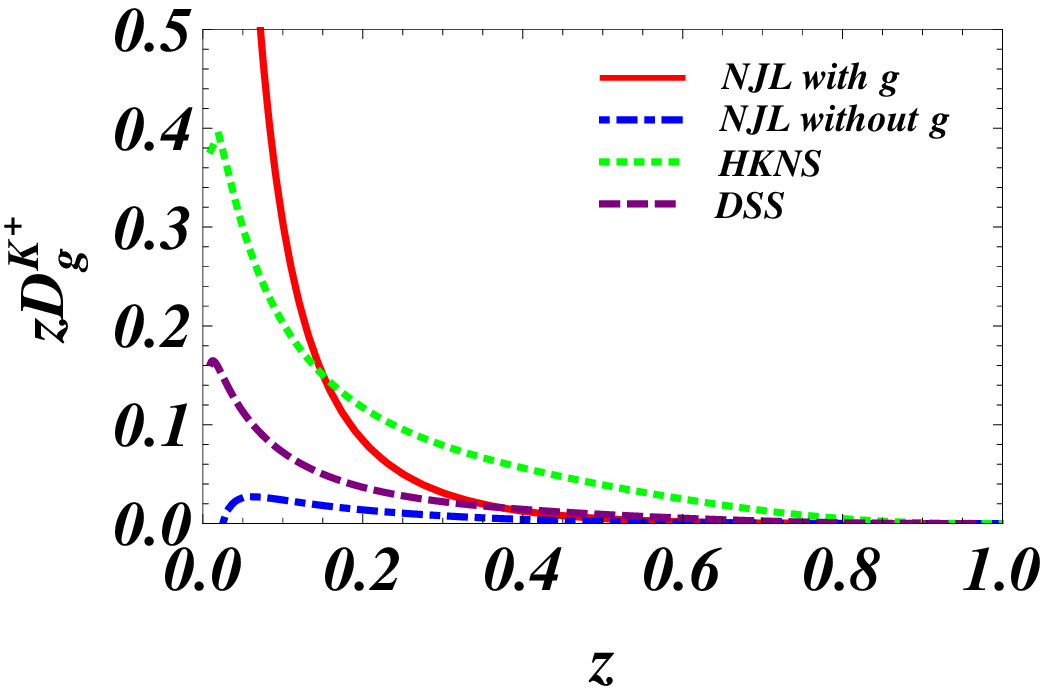}

\hspace{1.0cm}(c)\hspace{7.0cm}(d)
\caption{\label{fig37}
Same as Fig.~\ref{fig35}, but for the $K^+$ meson emission.}
\end{figure*}

Next we predict the $e^++e^-\rightarrow h+X$ differential cross section
\cite{ref42}
\begin{equation}
\begin{aligned}
F^h(z,Q^2)\equiv\frac{1}{\sigma_{tot}}\frac{d\sigma(e^++e^-\rightarrow h+X)}{dz},
\end{aligned}\label{diff}
\end{equation}
using the quark and gluon fragmentation functions from the previous section, where
$z=2E_h/\sqrt{s}=2E_h/Q$ is the energy fraction, with $E_h$ being
the energy carried by the hadron $h$, $\sqrt{s}$ being the center of mass energy, and
$Q$ being the invariant mass of the virtual photon or $Z$ boson. According to the
factorization theorem, Eq.~(\ref{diff}) can be written as
a convolution of two subprocesses \cite{ref41}: the hard scattering part
$e^++e^-\rightarrow\gamma (Z) \rightarrow q+\bar{q}$ at LO or
$e^++e^-\rightarrow\gamma (Z) \rightarrow q+\bar{q}+g$ at NLO, which is calculable
in perturbative QCD, and the hadronic part $q+\bar{q} (q+\bar{q}+g) \rightarrow h+X$,
which involves nonperturbative dynamics. The latter is described by the
fragmentation functions for the hadron $h$ emitted by the partons $q$, $\bar{q}$, or $g$.
We have the factorization formula \cite{ref42}
\begin{equation}
\begin{aligned}
F^h(z,Q^2)=\sum_i C_i(z,\alpha_s) \otimes D_i^h(z,Q^2),\label{f}
\end{aligned}
\end{equation}
where the subscript $i=u, d, s,...,g$ denotes flavors of partons, the coefficient
functions $C_i(z,\alpha_s)$ have been computed up to NLO in the modified minimal
subtraction scheme \cite{ref43}, and $D_i^h(z,Q^2)$ denotes the parton $i$
fragmentation function for the hadron $h$. The convolution $\otimes$ is defined by
\begin{equation}
\begin{aligned}
f(z) \otimes g(z)=\int_{z}^{1}\frac{dy}{y}f(y)g(\frac{z}{y}).
\end{aligned}
\end{equation}

\begin{figure*}
 \centering
\includegraphics[scale=0.55]{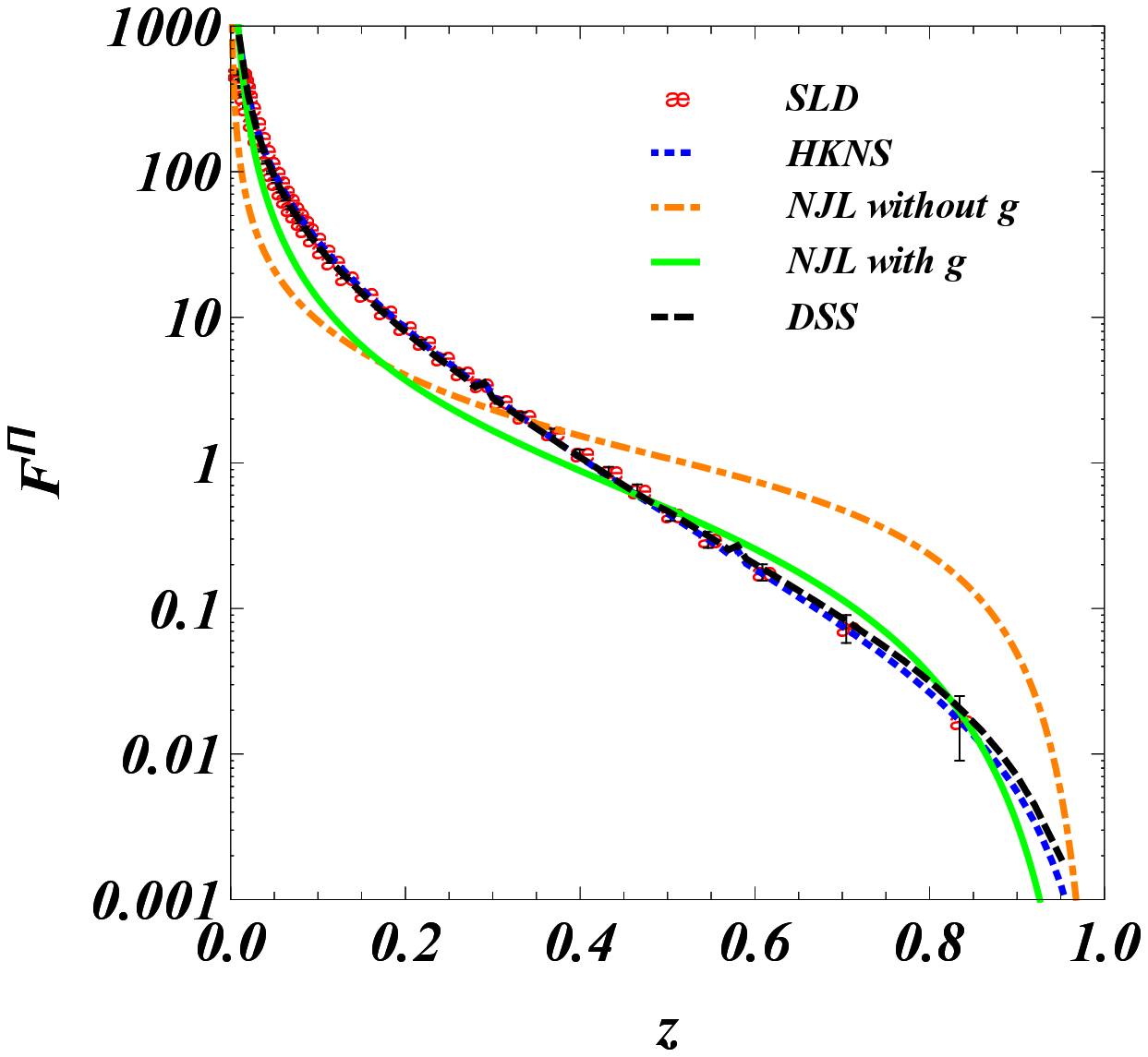}
\includegraphics[scale=0.55]{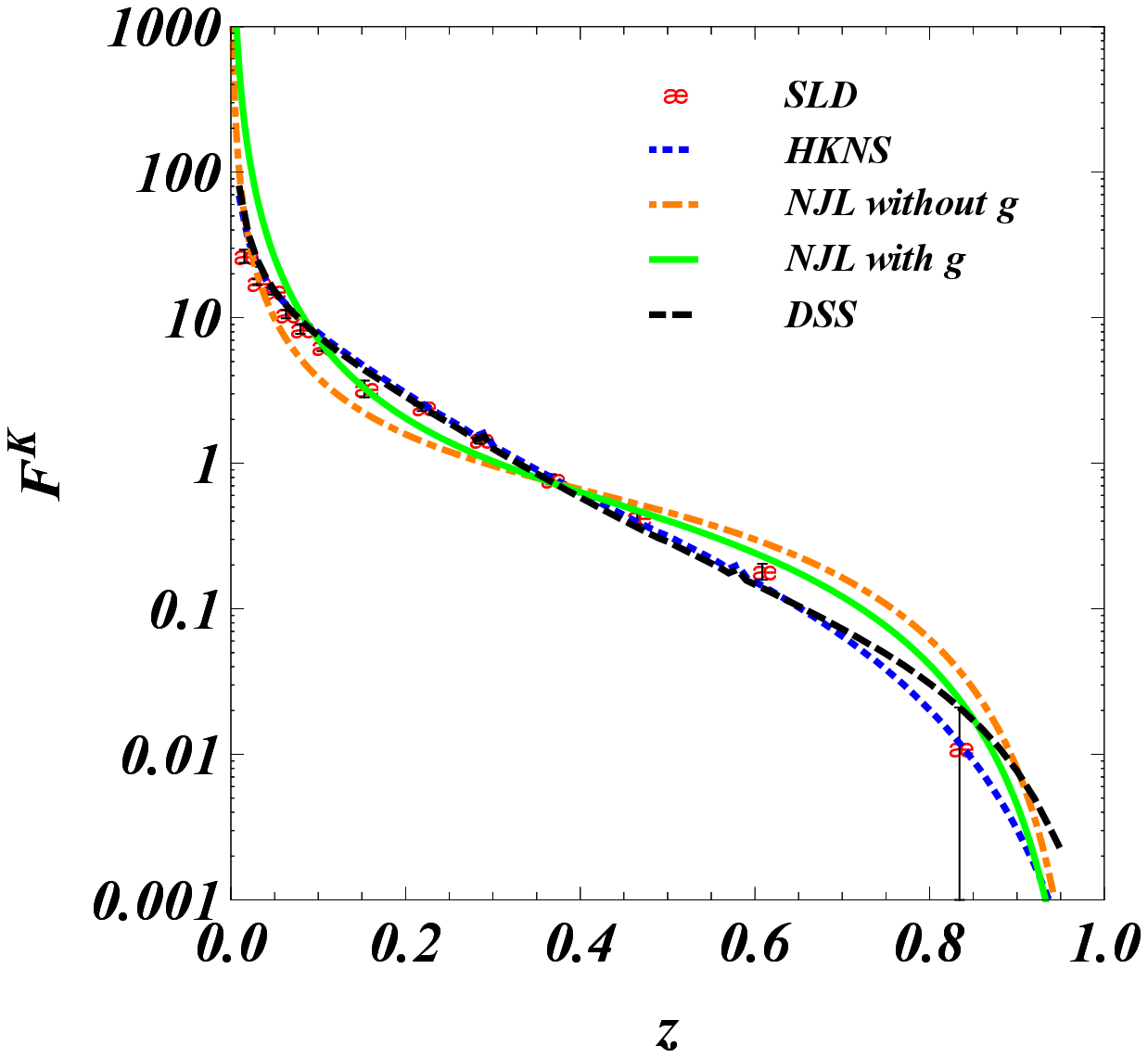}

\hspace{1.5cm}(a)\hspace{6.5cm}(b)

\includegraphics[scale=0.55]{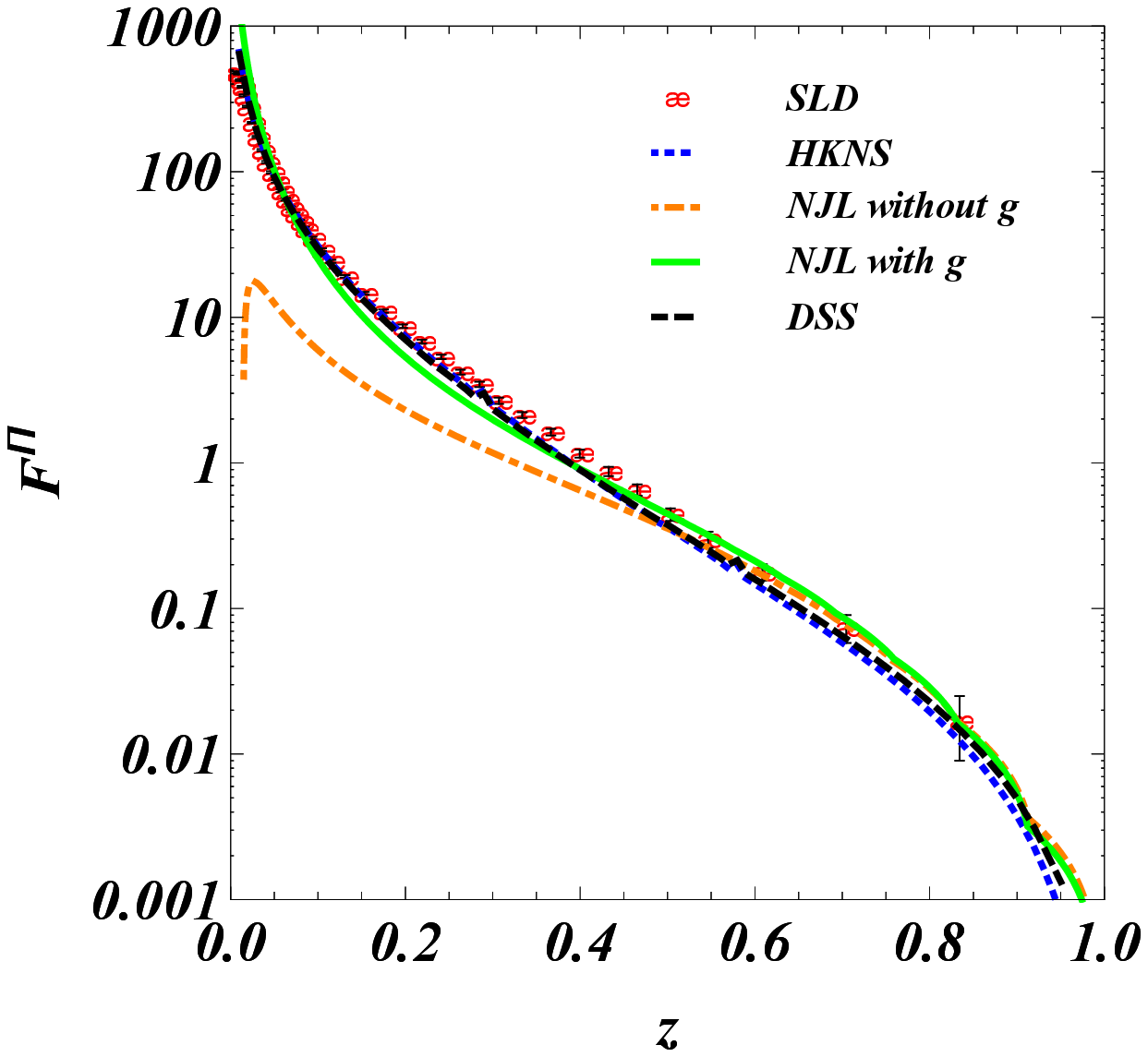}
\includegraphics[scale=0.55]{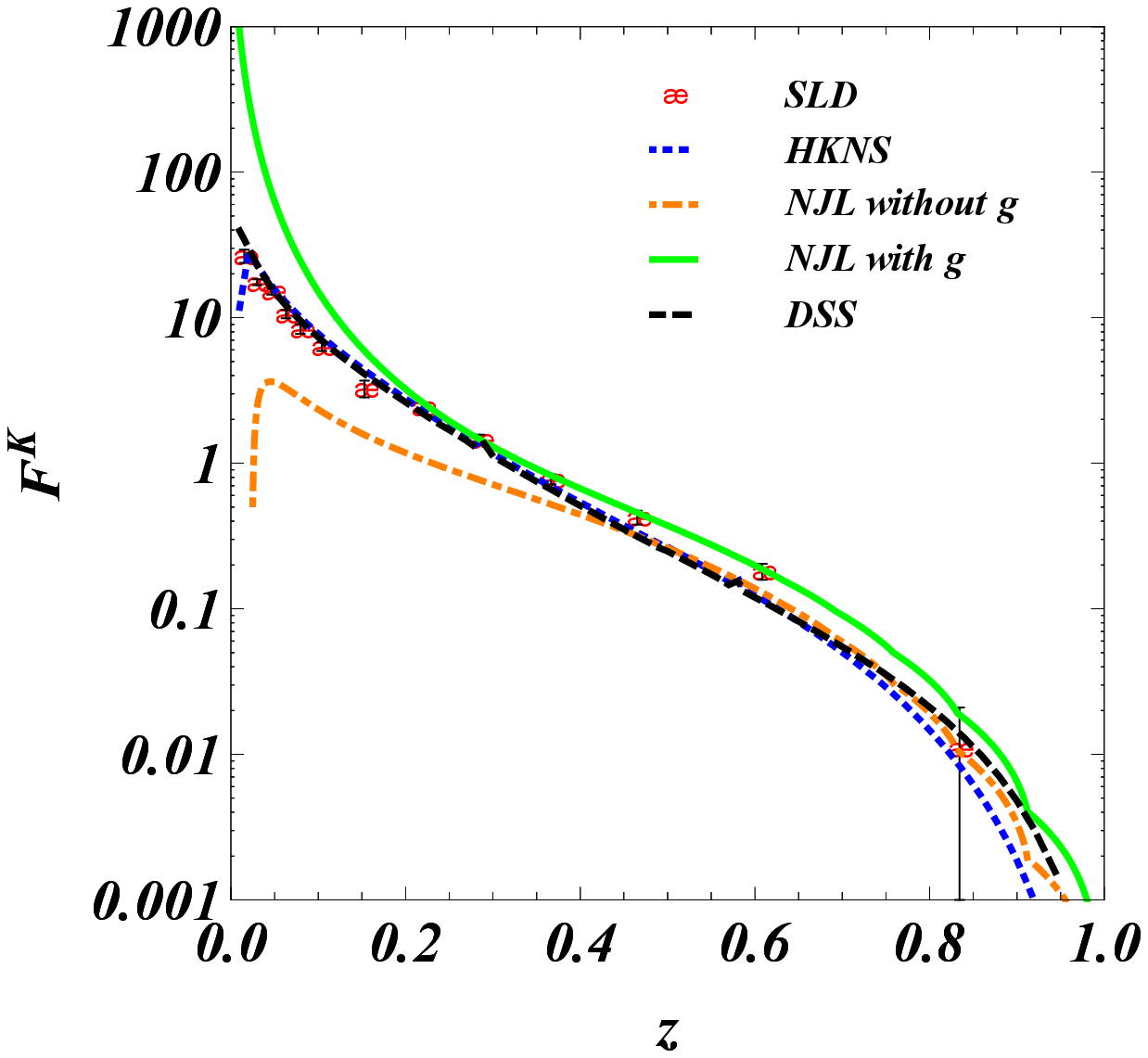}

\hspace{1.5cm}(c)\hspace{6.5cm}(d)
\caption{\label{fig39}
Our predictions for $F^h(z,Q^2)$ compared with the SLD data, and the HKNS and DSS
parameterizations at the scale $Q^2=M_Z^2$ for (a) $h=\pi$ and (b) $h=K$ under the
LO evolution. Same for (c) and (d) under the NLO evolution.}
\end{figure*}

Our predictions for $F^h(z,Q^2)$ in Eq.~(\ref{f}), $h=\pi$ and $K$, are compared
to the SLD data \cite{ref21} at the scale $Q^2=M_Z^2$ under the LO and NLO evolutions
in Fig.~\ref{fig39}. It is observed in all the plots that the curves labeled by
"NJL without g" are significantly lower than the SLD data for $z<0.4$, and
higher than the SLD data for $z>0.4$ in the pion channel. The inclusion of the
gluon fragmentation functions, correcting the above tendency, improves the overall
consistency with the data. This improvement highlights the phenomenological impact of
the gluon fragmentation functions, and their importance for accommodating the data.
In particular, the "NJL with g" predictions agree well with the SLD data in the pion
channel, after the NLO evolution is implemented. It is roughly the
case in the kaon channel, but with the "NJL with g" predictions overshooting the
data in the small $z<0.2$ region. However, the curves labeled by "NJL with g"
from the NLO evolution are very close to the HKNS and DSS parameterizations
in both the pion and kaon channels. The agreement of the predictions with the data
supports our proposal to treat a gluon as a pair of color lines in the NJL
model.

\begin{figure*}
 \centering
\includegraphics[scale=0.7]{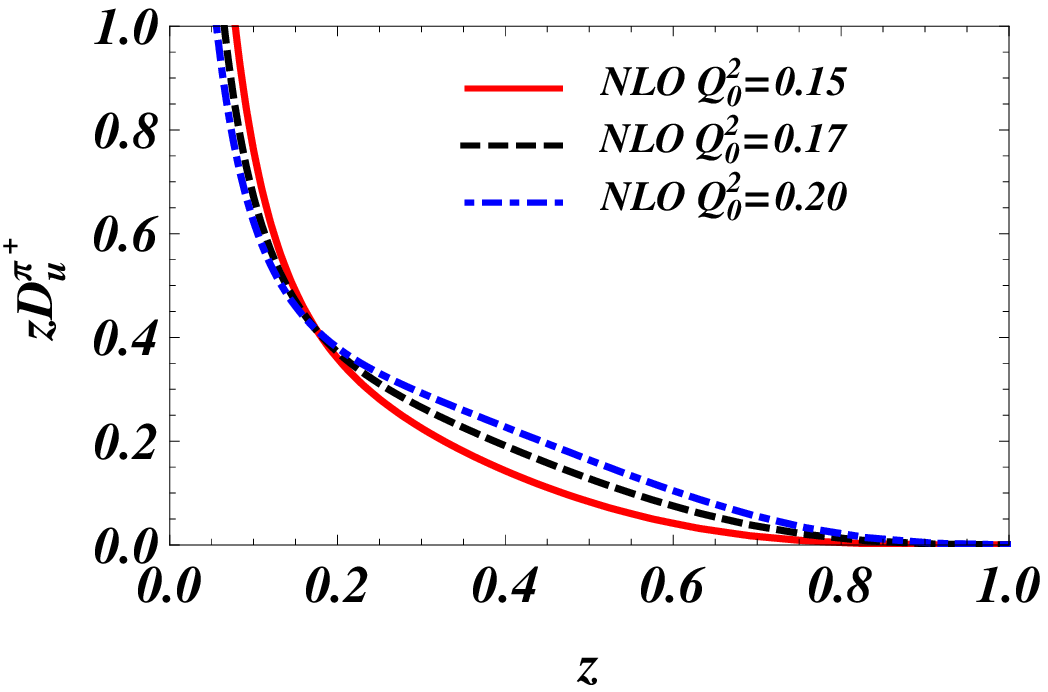}
\includegraphics[scale=0.7]{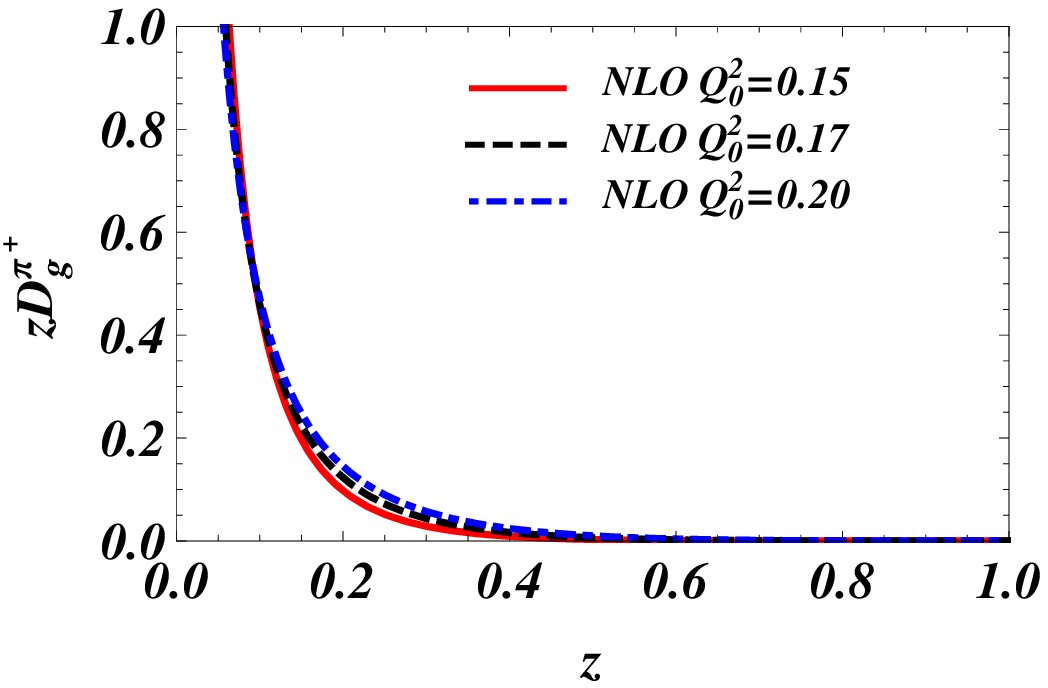}

\hspace{1.0cm}(a)\hspace{7.0cm}(b)
\caption{\label{fig41}
$z$ dependence of (a) $zD_{u}^{\pi^+}(z)$ and (b) $zD_{g}^{\pi^+}(z)$ at the scale
$Q^2=4$ GeV$^2$ for three different values of $Q_0^2$ (in units of GeV$^2$).}
\end{figure*}

\begin{figure*}
 \centering
\includegraphics[scale=0.7]{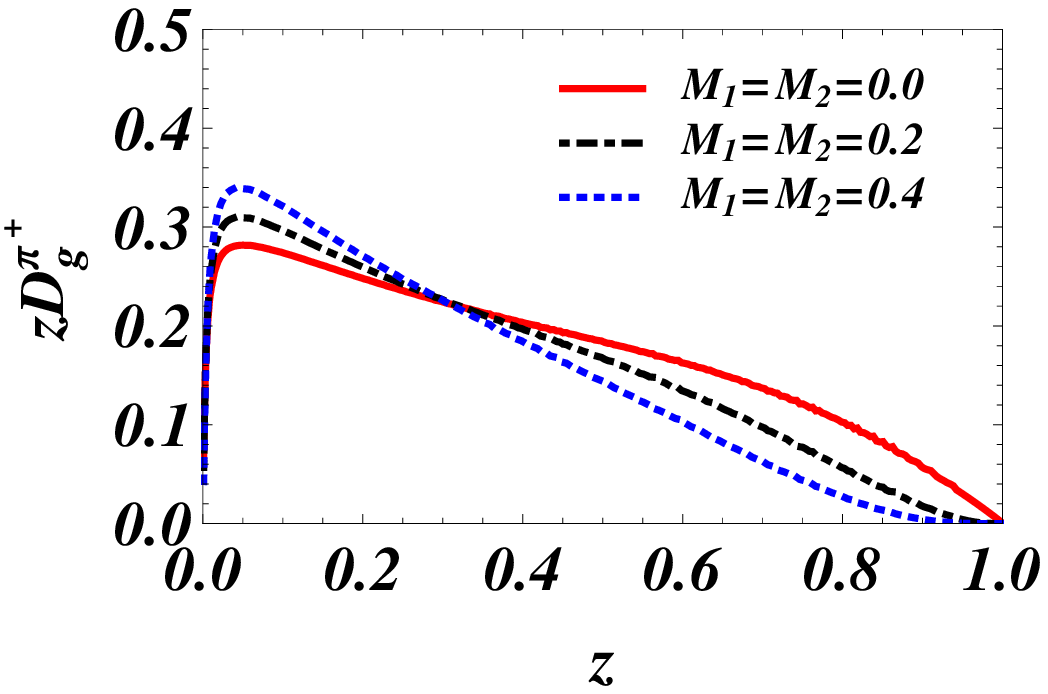}
\includegraphics[scale=0.7]{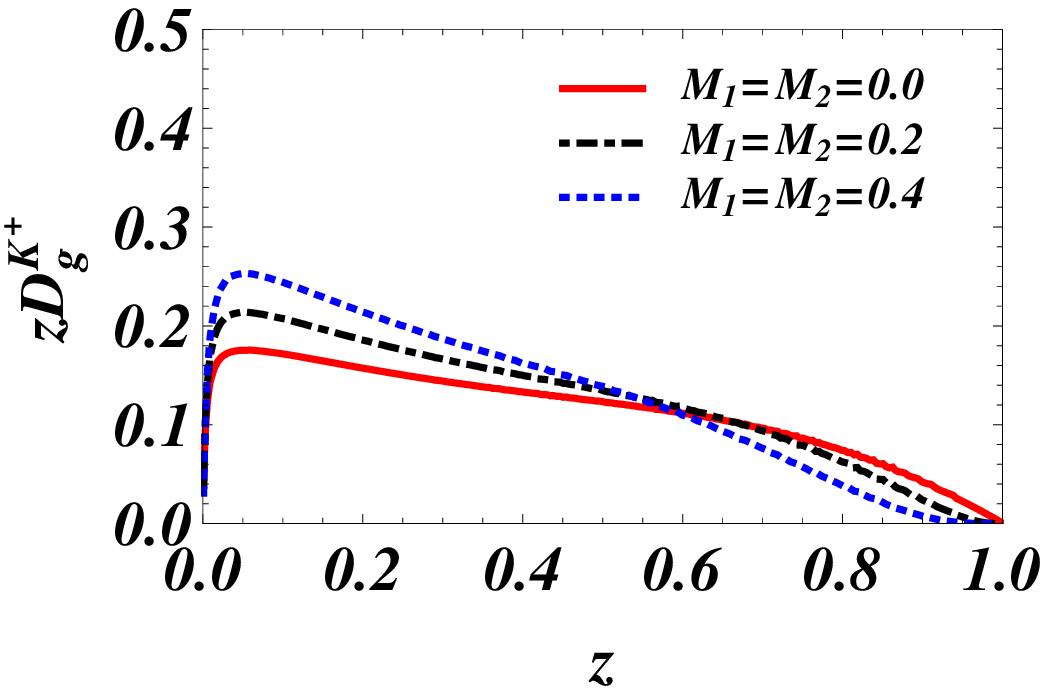}

\hspace{1.0cm}(a)\hspace{7.0cm}(b)
\caption{\label{fig43}
$z$ dependence of (a) $zD_{g}^{\pi^+}(z)$ and (b) $zD_{g}^{K^+}(z)$ at the model scale
for three different values of $M_1=M_2$ (in units of GeV).}
\end{figure*}

At last, we check the sensitivity of our results to the variation of some
model parameters. Figure~\ref{fig41} shows the $u$-quark and gluon fragmentation
functions for the $\pi^+$ meson under the NLO evolution from three different initial
model scales $Q_0^2=0.15$, 0.17, and 0.20 GeV$^2$ to $Q^2=4$ GeV$^2$. It is found
that the quark fragmentation function is more sensitive to the variation of the
model scale than the gluon fragmentation function. It hints that the
$e^++e^-\rightarrow h+X$ differential cross section at high $z$, dominated by the
contribution from the quark fragmentation functions, depends more strongly on the
model scale. We have taken into account this property, as determining the model
scales via reasonable fits of our predictions to the SLD data. The sensitivity of
the gluon fragmentation functions for the $\pi^+$ and $K^+$ mesons at the
model scale to the fictitious quark mass is examined in Fig.~\ref{fig43}. The
difference among the three sets of curves for $M_1=M_2=0.0$, 0.2, and 0.4 GeV in
the region of finite $z$ turns out to be easily smeared by the QCD evolution effect.
It explains why we have adopted the input $M_1=M_2=0.0$ for convenience in this work.

\section{CONCLUSION}

In this paper we have derived the gluon fragmentation functions in the NJL
model by treating a gluon as a pair of color lines formed by fictitious quark
and anti-quark under the requirement that they remain in the flavor-singlet
state after simultaneous meson emissions. The idea originates form the color
dipole model, in which the same treatment turns parton emissions into
emissions of color dipoles. The gluon fragmentation functions were then
formulated in terms of the quark fragmentation functions accordingly. The simplest
version of our proposal is consistent with that in the Lund model \cite{ref14},
as a combination of the quark and anti-quark fragmentation functions. A refined
version is to include the quark annihilation mechanism, such that the specific
flavor of the fictitious quarks is irrelevant, and the color lines just serve as
color sources of meson emissions. The corresponding gluon elementary fragmentation
functions constructed from the quark and anti-quark elementary fragmentation
functions lead to the integral equation, as a consequence of the iterations of
the elementary fragmentation into a jet process. The gluon branching effect, i.e.,
the multi-dipole contribution to the gluon fragmentation was also discussed in the
same framework, and found to be minor.

The results from the above three different schemes of handling subtle gluonic
dynamics were compared at the model scale, and evolved to higher scales.
It has been confirmed that the QCD evolution effect pushes the difference among the
three schemes to the region of very small $z$. This explains why our results are
stable with respect to the variation of the model parameters and to the choices of
the splitting functions. We have demonstrated that the inclusion of the gluon
fragmentation functions into the theoretical predictions from only the quark
fragmentation functions greatly improves the agreement with the SLD data for
the pion and kaon productions in $e^+e^-$ annihilation at the scale $Q^2=M_Z^2$.
Especially, our predictions for the pion emission from the NLO evolution are
well consistent with the SLD data, and with the HKNS and DSS parameterizations.
This nontrivial consistency implies that our proposal may have provided a plausible
construct for the gluon fragmentation functions, which are supposed to be null
in the NJL model.

The framework presented in this paper is ready for a generalization to the
quark and gluon fragmentation functions for other pseudoscalar mesons,
such as $\eta$ and $\eta'$. Wide applications are expected. The heavy-quark
(charm and bottom) fragmentation functions should be
included for a complete QCD evolution to $Q^2$ as high as $M_Z^2$, which
have been taken into account in the HKNS and DSS parameterizations. How to
establish the heavy-quark fragmentation functions in an effective model is
another challenging and important mission. We will address these subjects
in future works.

\appendix

\section{RESULTS UNDER LO EVOLUTION}

\begin{figure*}
 \centering
\includegraphics[scale=0.7]{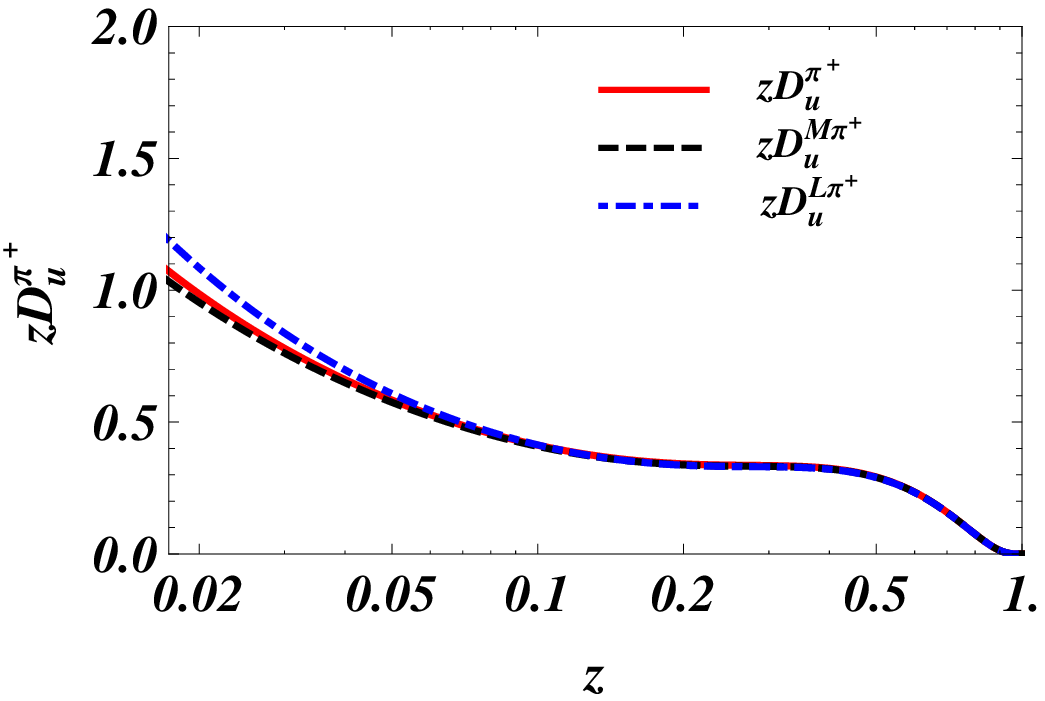}
\includegraphics[scale=0.7]{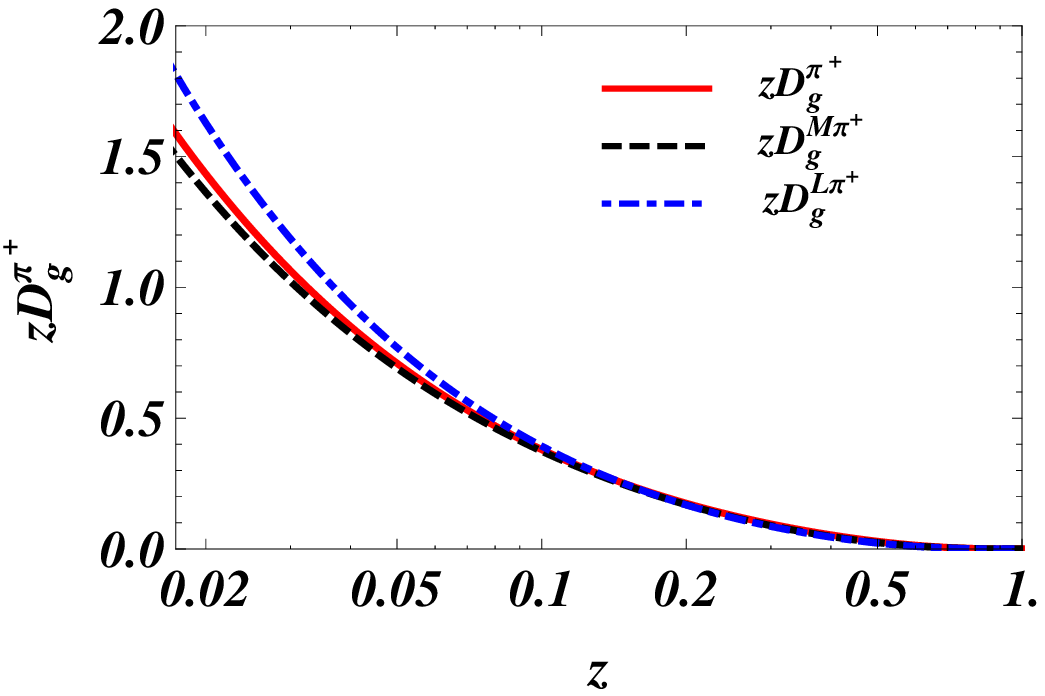}

\hspace{1.0cm}(a)\hspace{7.0cm}(b)
\includegraphics[scale=0.7]{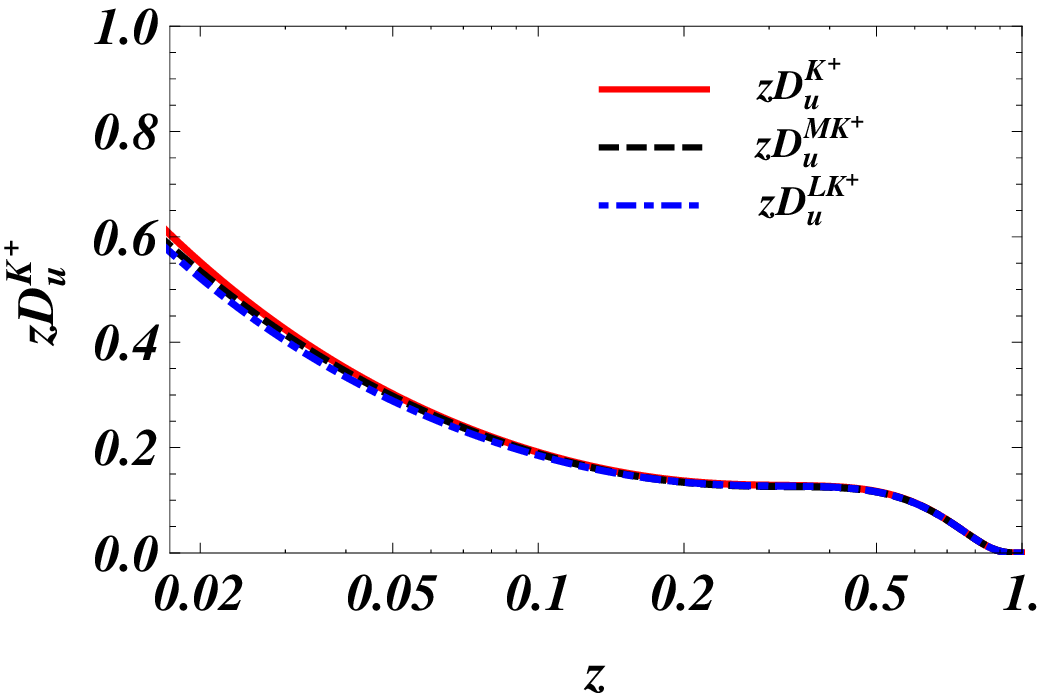}
\includegraphics[scale=0.7]{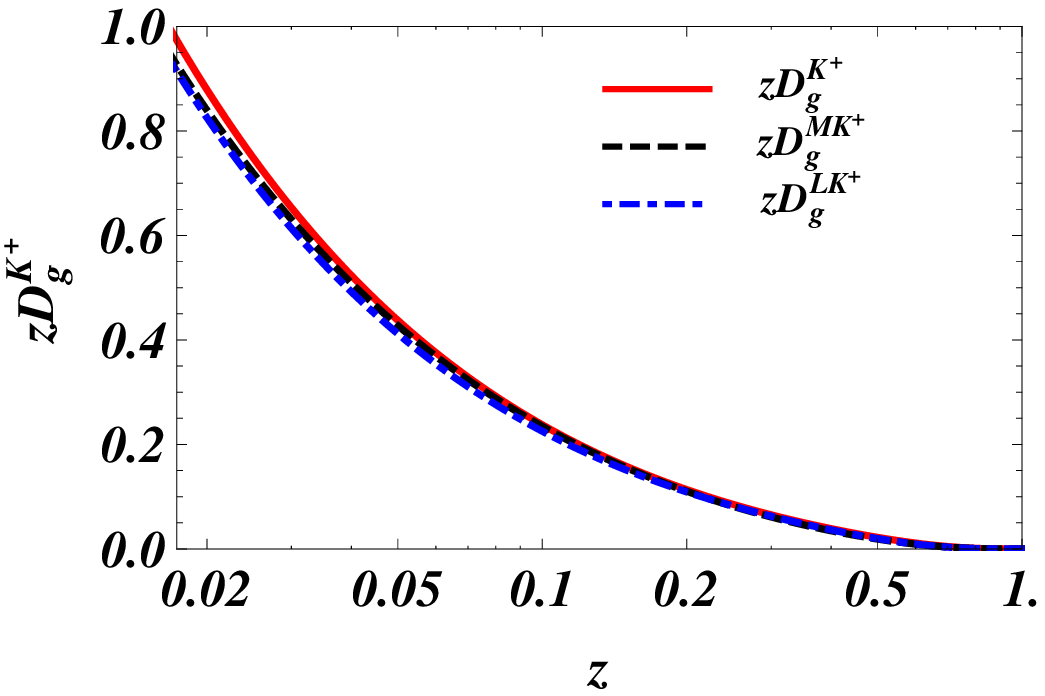}

\hspace{1.0cm}(c)\hspace{7.0cm}(d)
\caption{\label{fig19}
$z$ dependence of (a) $zD_{u}(z)$, $zD_{u}^{M}(z)$, and
$zD_{u}^{L}(z)$, and (b) $zD_{g}(z)$, $zD_{g}^{M}(z)$, and
$zD_{g}^{L}(z)$ for the $\pi^+$ meson emission at the scale
$Q^2=4$ GeV$^2$. (c) and (d) are for the $K^+$ meson emission.}
\end{figure*}

We collect some results from the LO evolution in this appendix.
Figure~\ref{fig19} displays the similarity of the quark and gluon
fragmentation functions from the the three different schemes, namely,
the scheme consistent with the Lund model, the scheme including
the quark annihilation mechanism, and the scheme including the
multi-dipole contribution, under the LO evolution. This similarity
supports the consideration of only the scheme with the quark
annihilation mechanism.

\begin{figure*}
 \centering
\includegraphics[scale=0.62]{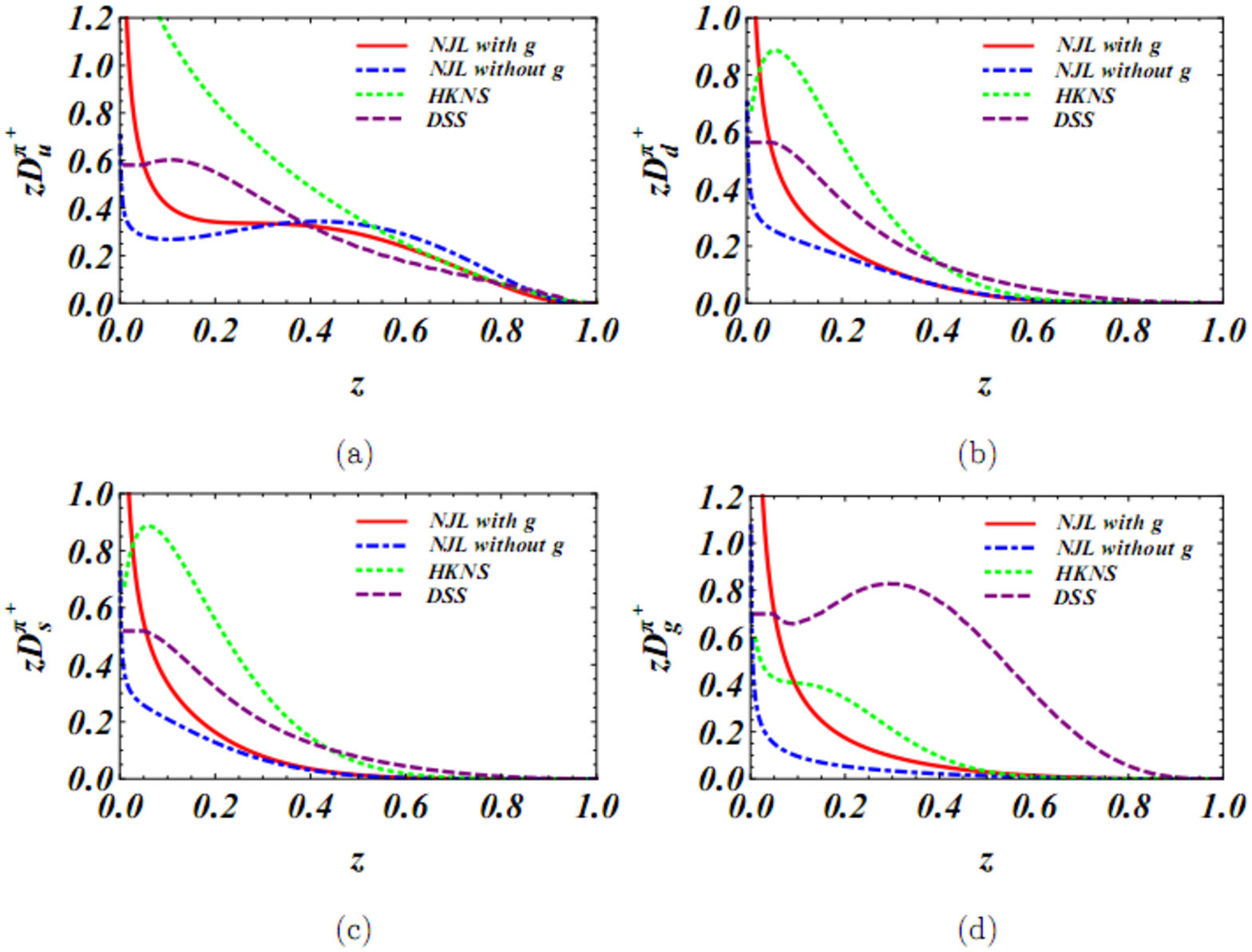}

\caption{\label{fig23}
Comparison of (a) $zD_{u}^{\pi^+}(z)$, (b) $zD_{d}^{\pi^+}(z)$, (c) $zD_{s}^{\pi^+}(z)$,
and (d) $zD_{g}^{\pi^+}(z)$ with the HKNS and DSS parameterizations at the scale
$Q^2=4$ GeV$^2$ under the LO evolution.}
\end{figure*}

\begin{figure*}
 \centering
\includegraphics[scale=0.62]{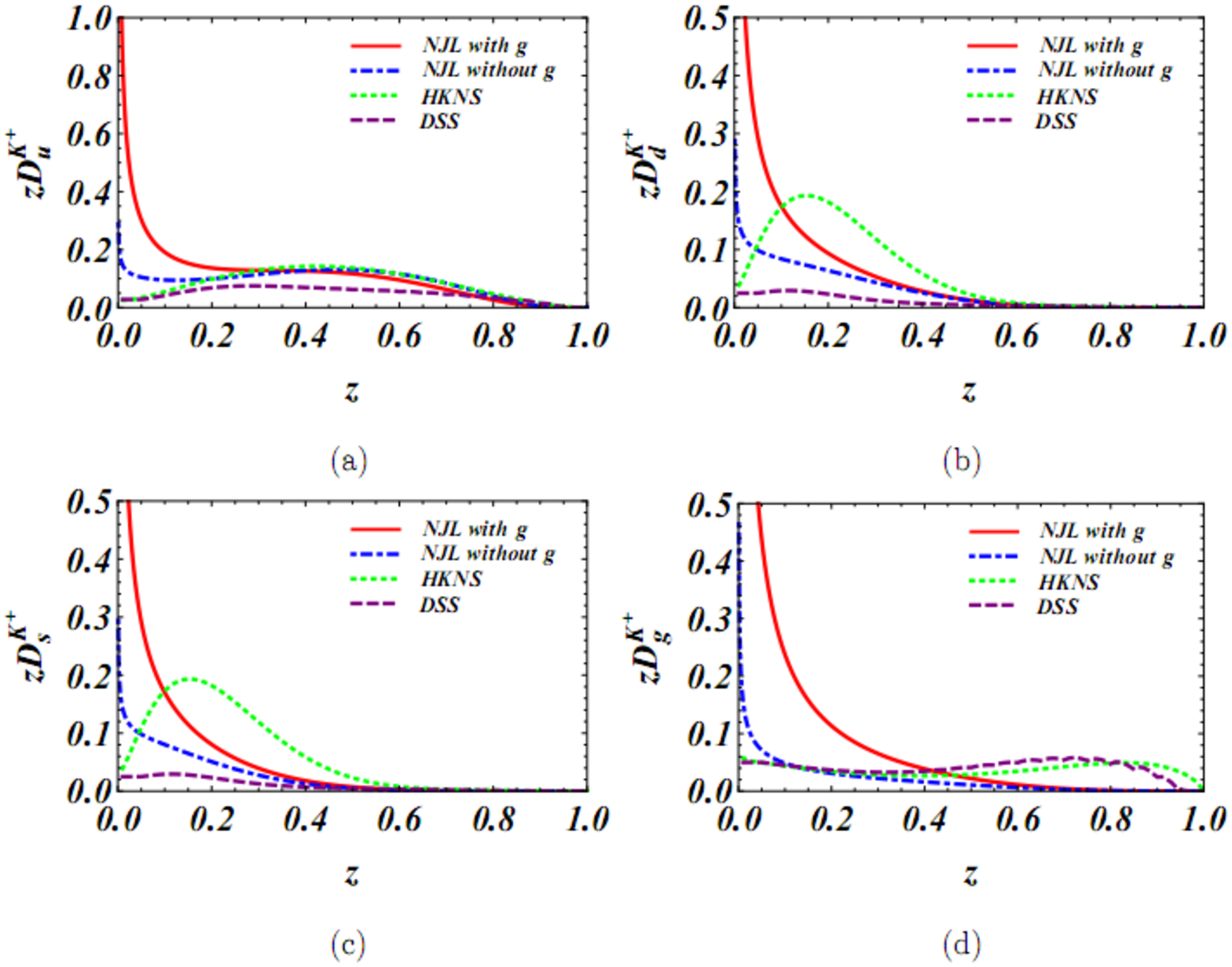}

\caption{\label{fig25}
Same as Fig.~\ref{fig23}, but for the $K^+$ meson emission.}
\end{figure*}

\begin{figure*}
 \centering
\includegraphics[scale=0.62]{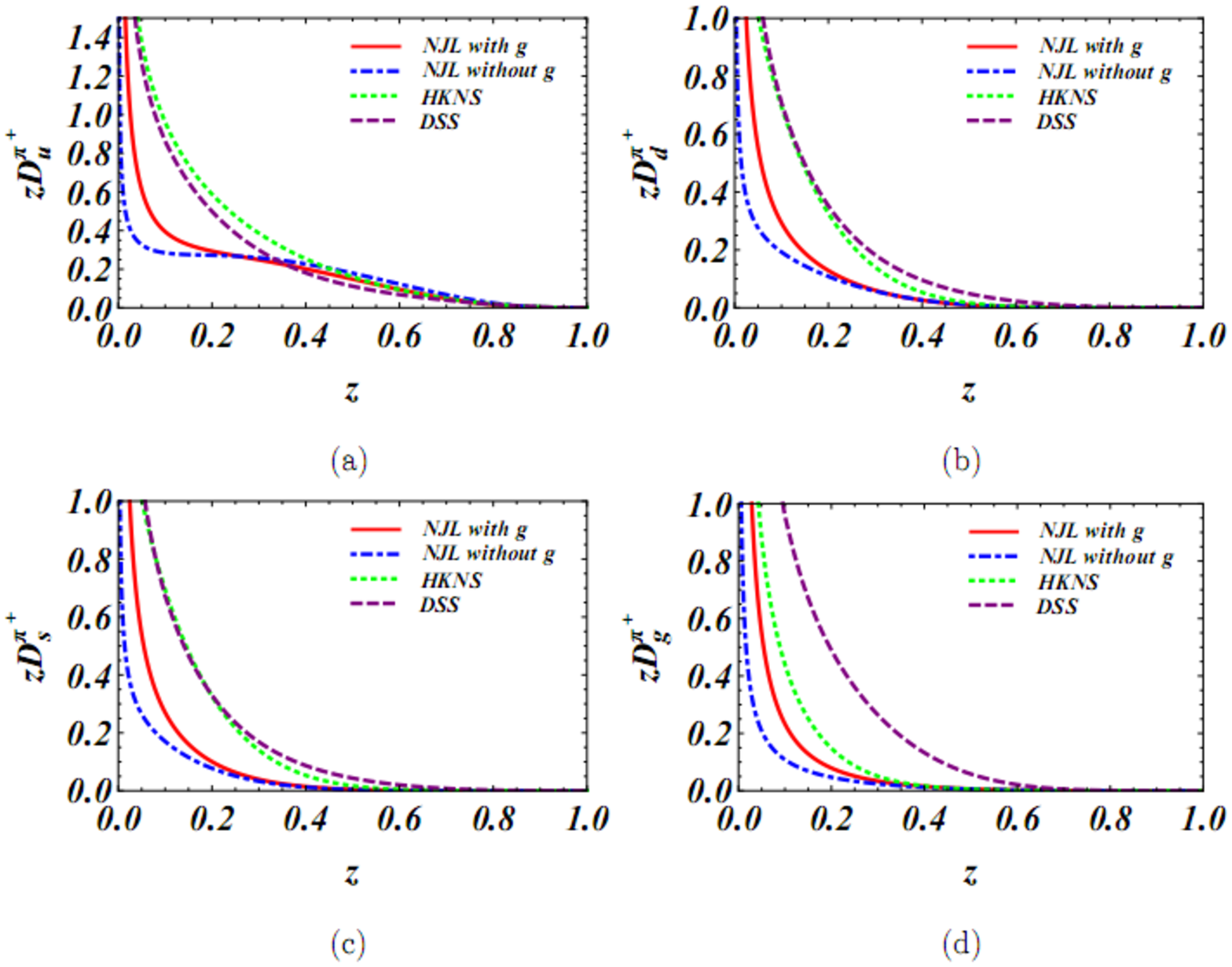}
\caption{\label{fig27}
Comparison of (a) $zD_{u}^{\pi^+}(z)$, (b) $zD_{d}^{\pi^+}(z)$, (c) $zD_{s}^{\pi^+}(z)$,
and (d) $zD_{g}^{\pi^+}(z)$ with the HKNS and DSS parameterizations at the scale
$Q^2=M_Z^2$ under the LO evolution.}
\end{figure*}

\begin{figure*}
 \centering
\includegraphics[scale=0.62]{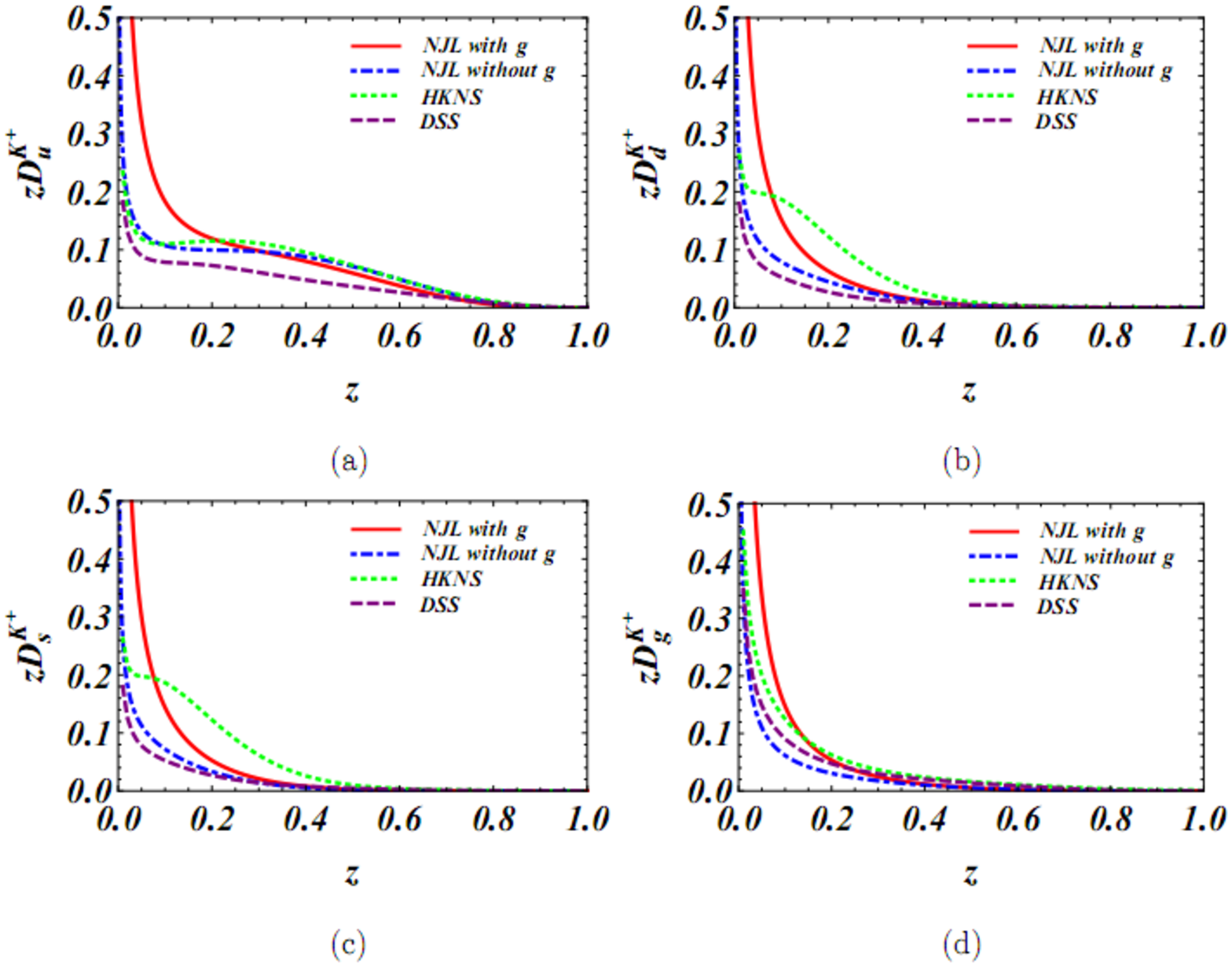}
\caption{\label{fig29}
Same as Fig.~\ref{fig27}, but for the $K^+$ meson emission.}
\end{figure*}

Our results for the quark and gluon fragmentation functions are compared with the
HKNS and DSS parameterizations at the scales $Q^2=4$ GeV$^2$ and $Q^2=M_Z^2$ under
the LO evolution in Figs.~\ref{fig23}-\ref{fig29}. Similar to the observation drawn
from the NLO evolution, the obvious difference between the curves labeled by
"NJL with g" and by "NJL without g" indicates the importance of the gluon
fragmentation functions. For any quark or gluon to the $\pi^+$ meson channels, the
"NJL with g" results agree better with the HKNS and DSS ones than the "NJL without g"
results do at $Q^2=4$ GeV$^2$ and $Q^2= M_Z^2$ in the almost entire region of $z$. For
the $K^+$ meson channels,  it is hard to tell which curves, "NJL with g" or
"NJL without g" are closer to the HKNS and DSS ones. Again, all the curves
are more distinct in the low $z$ region.

\begin{acknowledgments}

This work was supported  in part by the Ministry of Science and Technology of R.O.C..
Hsiang-nan Li is supported by Grant No. MOST-104-2112-M-001-037-MY3 from Ministry of Science and Technology (MOST) of Taiwan. Dong-Jing Yang is supported by MOST of Taiwan (Grant No. MOST-102-2112-M-033-005-MY3).
\end{acknowledgments}


\begin{thebibliography}{99}

\bibitem{ref1} J. C. Collins, Nucl. Phys. {\bf B396}, 161 (1993).
\bibitem{ref2} P. J. Mulders and R.D. Tangerman, Nucl. Phys. {\bf B461}, 197 (1996).
\bibitem{ref3} D. Boer and P. J. Mulders, Phys. Rev. D {\bf 57}, 5780 (1998).
\bibitem{ref4} M. Anselmino, M. Boglione, and F. Murgia, Phys. Lett. B {\bf 362}, 164 (1995).
\bibitem{ref5} M. Anselmino, M. Boglione, U. D'Alesio, E. Leader, S. Melis, F. Murgia,
and A. Prokudin, arXiv:0907.3999 [hep-ph].
\bibitem{ref6} E. Christova and E. Leader, Eur. Phys. J. C {\bf 51}, 825 (2007).
\bibitem{ref7} M. Anselmino, M. Boglione, U. D'Alesio, A. Kotzinian, F. Murgia,
A. Prokudin, and C. T\"urk, Phys. Rev. D {\bf 75}, 054032 (2007).
\bibitem{ref8} A. Bacchetta, M. Diehl, K. Goeke, A. Metz, P. J. Mulders, and M. Schlegel,
JHEP {\bf 02}, 093 (2007).
\bibitem{ref9} A.V. Efremov, K. Goeke, and P. Schweitzer, Phys. Rev. D {\bf 73}, 094025 (2006).
\bibitem{ref10} J. C. Collins, A.V. Efremov, K. Goeke, S. Menzel, A. Metz, and P. Schweitzer,
Phys. Rev. D {\bf 73}, 014021 (2006).
\bibitem{ref11} X. d. Ji, J. p. Ma, and F. Yuan, Phys. Rev. D {\bf 71}, 034005 (2005).
\bibitem{ref12} H. H. Matevosyan, A. W. Thomas, and W. Bentz, Phys. Rev. D {\bf 83}, 114010 (2011).
\bibitem{ref13} D. J. Yang, F. J. Jiang, C.W. Kao, and S. I. Nam, Phys. Rev. D {\bf 87}, 094007 (2013).
\bibitem{ref14} B. Andersson, G. Gustafson, G. Ingelman, and T. Sj\"ostrand, Phys. Rept. {\bf 97}, 31 (1983).
\bibitem{ref15} R. Brandelik et al. (TASSO Collaboration), Phys. Lett. B {\bf 94}, 444 (1980).
\bibitem{ref16} M. Althoff et al. (TASSO Collaboration), Z. Phys. C {\bf 17}, 5 (1983).
\bibitem{ref17} W. Braunschweig et al. (TASSO Collaboration), Z. Phys. C {\bf 42}, 189 (1989).
\bibitem{ref18} H. Aihara et al. (TPC Collaboration), Phys. Rev. Lett. {\bf 52}, 577 (1984);
{\bf 61}, 1263 (1988).
\bibitem{ref19} M. Derrick et al. (HRS Collaboration), Phys. Rev. D {\bf 35}, 2639 (1987).
\bibitem{ref20} R. Itoh et al. (TOPAZ Collaboration), Phys. Lett. B {\bf 345}, 335 (1995).
\bibitem{ref21} K. Abe et al. (SLD Collaboration), Phys. Rev. {\bf D69}, 072003 (2004).
\bibitem{ref22} D. Buskulic et al. (ALEPH Collaboration), Z. Phys. C {\bf 66}, 355 (1995);
R. Barate et al., Phys. Rep. {\bf 294}, 1 (1998).
\bibitem{ref23} R. Akers et al. (OPAL Collaboration), Z. Phys. C {\bf 63}, 181 (1994).
\bibitem{ref24} P. Abreu et al. (DELPHI Collaboration), Eur. Phys. J. {\bf C5}, 585 (1998).
\bibitem{ref25} P. Abreu et al. (DELPHI Collaboration), Nucl. Phys. {\bf B444}, 3 (1995).
\bibitem{ref26} M. Hirai, S. Kumano, T. H. Nagai, and K. Sudoh, Phys. Rev. D {\bf 75}, 094009 (2007).
\bibitem{ref27} D. de Florian, R. Sassot, and M. Stratmann, Phys. Rev. D {\bf 75}, 114010 (2007).
\bibitem{ref28} Y. Nambu, G. Jona-Lasinio, Phys. Rev. {\bf 122}, 345 (1961).
\bibitem{ref29} Y. Nambu, G. Jona-Lasinio, Phys. Rev. {\bf 124}, 246 (1961).
\bibitem{ref30} T. Shigetani, K. Suzuki and H. Toki,
Phys. Lett. B {\bf 308}, 383 (1993).  
\bibitem{ref31} R. M. Davidson and E. Ruiz Arriola, Phys. Lett. B {\bf 359}, 273 (1995).
\bibitem{ref32} H. H. Matevosyan, A. W. Thomas, and W. Bentz, Phys. Rev. D {\bf 83}, 074003 (2011).
\bibitem{ref33} H. H. Matevosyan, W. Bentz, I. C. Cloet, and A. W. Thomas,
Phys. Rev. D {\bf 85}, 014021 (2012).
\bibitem{ref34} R. D. Field and R. P. Feynman, Nucl. Phys. {\bf B136}, 1 (1978).
\bibitem{ref35} G. Gustafson, Phys. Lett. B {\bf 175}, 453 (1986).
\bibitem{ref36} G. Gustafson and U. Pettersson, Nucl. Phys. {\bf B306}, 746 (1988).
\bibitem{ref37} L. B. Andersson, G. Gustafson, L\"onnblad, and U. Pettersson, Z. Phys. C
{\bf 43}, 625 (1989).
\bibitem{ref38} A. H. Mueller and B. Patel, Nucl. Phys. {\bf B425}, 471 (1994).
\bibitem{ref39} V.N. Gribov and L.N. Lipatov, Sov. J. Nucl. Phys. {\bf 15}, 428 (1972);
G. Altarelli and G. Parisi, Nucl. Phys. {\bf B126}, 298 (1977);
Yu.L. Dokshitzer, Sov. Phys. JETP {\bf 46}, 641 (1977).
\bibitem{ref40} M. Botje, Comput. Phys. Commun. {\bf 182}, 490 (2011).
\bibitem{ref41} F. Halzen and A. D. Martin, {\it Quarks and Leptons: An Introductory
Course in Modern Particle Physics}, John Wiley \& Sons (1984).
\bibitem{ref42} R. K. Ellis, W. J. Stirling, and B. R. Webber, {\it QCD and Collider Physics},
Cambridge University Press (1996).
\bibitem{ref43} S. Kretzer, Phys. Rev. D {\bf 62}, 054001 (2000).

\end{thebibliography}
\end{document}